\documentclass[10pt]{iopart}

\usepackage{amssymb,amsfonts}
\usepackage{iopams}    
\usepackage{graphicx}  
\usepackage{bm}        

\usepackage{doi}
\usepackage{hyperref}

\usepackage{subcaption}
\usepackage{caption}
\usepackage{graphicx}
\usepackage{float} 

\usepackage{amssymb,amsfonts}

\usepackage[table]{xcolor}

\newcommand{\text}[1]{{\mbox{#1}}}
\newcommand{\eqref}[1]{(\ref{#1})}
\newcommand{\boxed}[1]{\fbox{$\displaystyle #1$}}
\newcommand{\casescustom}[1]{%
\left\{
\begin{array}{ll}
#1
\end{array}
\right.
}

\newcommand{\myfl}{\fl\qquad}

\def\dd{{\rm d}} 
\def\ii{{\rm i}} 

\def\Dlap{{\bm \nabla}^2}
\def\myL{L}
\def\flat{{\rm fl}}

\newcommand{\gr}[1]{{\bm #1}}
\def\affine{s} 
\def\alphaG{\xi}%
\newcommand{\mycomment}[1]{{}}

\definecolor{rossos}{cmyk}{0,1,1,0.55}
\definecolor{blu}{cmyk}{1,1,0,0.3}
\definecolor{bluc}{cmyk}{1,1,0,0.1}
\definecolor{verde}{cmyk}{0.92,0,0.59,0.25}
\definecolor{verdec}{cmyk}{0.92,0,0.59,0.15}
\definecolor{verdes}{cmyk}{0.92,0,0.59,0.4}
\hypersetup{colorlinks, bookmarksopen, bookmarksnumbered,citecolor=blu, linkcolor=bluc, pdfstartview=FitH, urlcolor=rossos}

\usepackage{doi}

\begin{document}


\title{Gravitational lensing beyond the eikonal approximation}
\author{Emma Bruyère$^1$ and Cyril Pitrou$^1$}

\address{$^1$Institut d'Astrophysique de Paris, UMR-7095 du CNRS et de Sorbonne Universit\'e, Paris, France}

\ead{bruyere@iap.fr}

\date{\today}

\begin{abstract}
Waves propagating through a gravitational potential exhibit wave-optics effects when their wavelength is not significantly smaller than the lensing scales. We study the propagation of a scalar wave, governed by the Klein-Gordon equation in curved spacetime, to focus on effects on amplitude and phase, while leaving aside the issue of wave polarization which affects electromagnetic and gravitational waves. Using the Newman-Penrose formalism, we obtain the first corrections beyond the geometric optics in the expansion in the inverse frequency. In vacuum, that is for Weyl tensor lensing, there is no wave effect at first order in $G$ and wave effects start at order $G^2$. Conversely, if the wave travels through a non-vanishing matter density, the first corrections start at order $G$. We check these analytic results by solving numerically the equations dictating the evolution of the corrections either in the vicinity of a Schwarzschild black hole or through a transparent star.
\end{abstract}

\section{Introduction}

Geometric optics, also called the eikonal approximation for wave propagation, is valid in the limit of large frequencies $\omega$, or equivalently for small wavelengths compared to the typical lengths of the physical system. It is equivalent to a classical particle description whose trajectories are the rays, and whose normal to these trajectories are the wavefront, hence it reduces to the Hamilton-Jacobi formulation of mechanics. Rays are null geodesics in the case of a massless field, and transport equations dictate how the wave amplitude and its polarization evolve along them~\cite{Isaacson:1967zz,SEF1992,Bartelmann:2010fz,Cusin:2024git}.

The eikonal approximation is expected to hold in most optical gravitational lensing scenarios due to the large scale separation between the light wavelength and the curvature or lensing scales. Lensing effects signatures are also investigated in current or forthcoming data~\cite{Sereno:2010dr,Jung:2017flg,Lai:2018rto,Oguri:2018muv,Dai:2018enj,Diego:2019lcd,Hannuksela:2019kle,Liu:2020par,Urrutia:2021qak,LIGOScientific:2021izm,LIGOScientific:2023bwz,Yin:2023kzr}. However, gravitational waves are typically produced with macroscopic wavelengths, and in some situations wave effects must be taken into account~\cite{LIGOScientific:2025rsn}. The physical effects are very similar to those of Fourier optics, and it has been established that the diffraction integral~\cite{Bliokh1975,Bontz:1981rvr,1985A&A...148..369S,Deguchi:1986zz,1986ApJ...307...30D,SEF1992,Ulmer:1994ij,Jaroszynski:1995cd,Nakamura:1999uwi,Takahashi:2004mc,Macquart:2004sh,universe9050200}, which is a suitable approximation of the Kirchhoff integral~\cite{Takahashi:2003ix,Takahashi:2005sxa}, describes the dominant wave effects. The recent detection of gravitational waves at low frequency with pulsar timing arrays~\cite{NANOGrav:2023gor,NANOGrav:2024tnd} and the forthcoming LISA experiment~\cite{LISA:2022yao} have sparked a renewed interest in the theoretical foundations of this approach~\cite{Takahashi:2016jom,Cusin:2019rmt,Dalang:2021qhu,Feldbrugge:2020ycp,Oancea:2020khc,Oancea:2022szu,Kubota:2023dlz,Leung:2023lmq,Jow:2022pux}, and it has been shown that it can also be formulated as a scattering problem~\cite{Pijnenburg:2024btj,Braga:2024pik,CarrilloGonzalez:2025gqm}. 

Instead of devising a method to compute the diffraction integral in full generality, e.g. numerically~\cite{Villarrubia-Rojo:2024xcj}, we aim here to compute the corrections to the eikonal description in powers of $1/\omega$. 
Following~\cite{Dolan:2017zgu,Dolan:2018ydp,Harte:2018wni,Harte:2019tid}, we use the Newman-Penrose formalism~\cite{Newman:1961qr}, in which the behaviour of a bundle of geodesics is dictated by a set of coupled ordinary differential equations in the affine parameter of a reference geodesic. The number of variables to solve depends on the order in $1/\omega$ required and it extends the usual geodesic deviation equation formulated in terms of deformation rate~\cite{Perlick:2010zh,Fleury:2015hgz}. The beyond-eikonal (BE) corrections of the wave amplitude are also obtained by solving a set of coupled transport equations for the wave and its successive gradient components. Although computationally involved, this method is systematic and theoretical expressions for the set of equations can be obtained thanks to a computer algebra system such as xAct~\cite{xAct}. 

Since BE effects would alter both the amplitude, the phase and the polarization of the wave, a first step when evaluating these effects consists in studying the case of a scalar wave whose propagation is dictated by the massless Klein-Gordon equation on a curved background, so as to focus on the effect on the amplitude and the phase only. This formalism is developed in section~\eqref{SecFormalism}. We then specify to the case of vacuum lensing in section~\eqref{SecWeyl}, and we find that for a pure Weyl-tensor-induced lensing, there is no BE effect at linear order in $G$ and corrections start at order $G^2$. In contrast, in section~\eqref{SecRicci} we detail the case of Ricci dominated lensing due to matter density. If the wave travels through a non-vanishing matter density in the lens plane, BE effects start at order $G$. In the thin-lens and weak-field approximation we are able to find accurate analytic approximations. We check the Weyl-lensing case by solving numerically the BE set of differential equations for lensing around a Schwarzschild black hole, and the Ricci-lensing case by considering a source placed exactly behind a spherical star described by a toy-model Tolman-Oppenheimer-Volkoff (TOV) solution~\cite{OppenheimerVolkoff1939,Tolman1939,Chan:2014tva}. Finally we conclude by discussing our differences with previous literature in section~\ref{SecConlusion}. 

We set $c=1$ throughout but we keep track of $G$ explicitly. We use $a,b,c,\dots$ for abstract spacetime tensorial indices and $i,j,k,\dots$ for their spatial counterparts.

\section{Beyond eikonal formalism}\label{SecFormalism}

\subsection{Eikonal expansion}

We consider a real scalar field $\Psi$ propagating on a background metric which describes the geometric structure of the lens. 
The wave evolution is dictated by the massless Klein-Gordon (KG) equation
\begin{equation}\label{waveq}
\square\Psi=0\,.
\end{equation}
For convenience we introduce a complex field $\psi$  such that $\Psi=\text{Re}(\psi)$~\cite{Harte:2018wni} which satisfies the same equation. The eikonal expansion is similar to a WKB expansion, as we assume that this complex field is the product of a slowly varying amplitude and a fast varying phase, that is
\begin{equation}\label{DefAphi}
\psi = A {\rm e}^{\ii \omega \varphi}\,.
\end{equation}
We then define the (dimensionless) wave vector by 
\begin{equation}
k_a \equiv \nabla_a\varphi\,.
\end{equation}
The convention chosen is such that $k^0 = 1$ in flat spacetime, that is $\omega \varphi = \omega (\gr{n}\cdot\gr{x}-t)$, as we also consider a mostly positive metric. Replacing the ansatz~\eqref{DefAphi} in~\eqref{waveq} we obtain the equivalent formulation
\begin{equation}
-\omega^2 k_a k^a A+2 \ii \omega k^a\nabla_a A+\ii \omega A \nabla_a k^a + \square A=0\,.
\end{equation}
It is generally assumed that the various powers of $\omega$ must vanish separately in the previous expression.  However this statement would only be valid if the previous expression was valid for all or several values of $\omega$ whereas here it takes only one value for a given wave.

Instead, we {\it assume} that the wavevector is null, which is always possible since~\eqref{DefAphi} is an arbitrary decomposition. This is of course motivated by the fact that it is increasingly accurate that the amplitude is slowly varying for large $\omega$. We then obtain that the rays are null geodesics and a transport equation for the amplitude, that is 
\begin{equation}\label{GeneralEikonalSplitting}
\fl \qquad k^a k_a=0\,,\quad \nabla_{[a} k_{b]}=0 \quad \Rightarrow\quad 
\casescustom{
  k^a \nabla_a k_b=0\,, \\
 2k^a\nabla_a A+A \nabla_a k^a= \ii \omega^{-1} \square A\,.
 }
\end{equation}
So far the formulation is totally equivalent to the KG equation. However, if the frequency is large with respect to gradients, this last equation can be solved perturbatively. We build the solution of $A$ iteratively as~\cite{MTW}
\begin{equation}\label{DecAAA}
A = A_0+\omega^{-1}A_1+ \omega^{-2}A_2 + \cdots \,\,,
\end{equation}
where typically $A_n/\omega^n \ll A_{n-1}/\omega^{n-1}$. The contribution at each order depends on the previous order solution since replacement in~\eqref{GeneralEikonalSplitting} leads to the coupled set
\begin{eqnarray}
& 2k^a\nabla_a A_0+A_0\nabla_a k^a=0 \label{eqA0} \,, \\
& 2k^a\nabla_a A_n+A_n\nabla_a k^a=\ii\square A_{n-1} \,.
\end{eqnarray}
This last transport equation can be rephrased as a transport equation along a null ray for the relative correction $A_n/A_0$, namely
\begin{equation}\label{TransportAn}
k^a\nabla_a\left(\frac{A_n}{A_0}\right)=\frac{\ii}{2}\frac{\square A_{n-1}}{A_0} \,.
\end{equation}
This is the central equation that we wish to solve recursively along a null geodesics so as to reconstruct the full solution~\eqref{DecAAA}. We can always choose $A_0$ real, therefore, we see from the above equation that corrections with $n$ odd will be purely imaginary and the ones with $n$ even will be purely real.

This procedure is however not straightforward because we need to know $\square A_{n-1}$ and not simply $A_{n-1}$ when we solve for the transport of $A_n$ along the null ray. In general solving for $A_n$ requires the knowledge on the null geodesics of the terms $\nabla_{a_1}\cdots\nabla_{a_{p}}A_0$ with $p\leq 2n$.

\subsection{Amplification factor}

If we keep only the first corrections to the amplitude and to the phase, we can separate the magnitude and phase in~\eqref{DefAphi} as~\cite{Dalang:2021qhu}
\begin{equation}
\fl\qquad \psi \simeq A_0\left[1+\omega^{-2} \left(\frac{\Im(A_1)^2}{2A_0^2} +\frac{A_2}{A_0}\right) \right] \exp\left(\ii\omega\varphi + \ii\omega^{-1}\frac{\Im(A_1)}{A_0}\right)\,.
\end{equation}
Thus, the first BE correction, $A_1$, modifies the phase. However it cannot be factorized as a frequency-independent time delay due to the $\omega^{-1}$ prefactor, and it will therefore alter the shape of a wave packet. Hence even in a situation of weak lensing with a single image of the source, that is without interference of several images in the geometric optics, this BE effect will alter non monochromatic signals.

Gravitational lensing is defined with respect to a reference situation where no gravitational effects occur. It is customary to define an amplification factor $F$ by
\begin{equation}
F \equiv \frac{A^{\rm lens}\,{\rm e}^{\ii \omega \varphi_{\rm lens}}}{A_0^{\rm no\,lens}\,{\rm e}^{\ii \omega \varphi_{\rm no\,lens}}}=F_{\rm GO} F_{\rm BE}\,,
\end{equation}
where the last equality defines a separation into an eikonal, or geometric optics (GO), contribution, and a BE contribution. The GO contribution is simply $F_{\rm GO} = A_0^{\rm lens}/A_0^{\rm no\,lens} {\rm e}^{\ii \omega (\varphi_{\rm lens}-\varphi_{\rm no\,lens})}$ and the BE amplification factor is $F_{\rm BE}=A^{\rm lens}/A^{\rm lens}_0$. Note that by construction we consider the same null geodesic when solving the transport equation~\eqref{TransportAn}, hence $\varphi_{\rm lens}-\varphi_{\rm no\,lens}$ does not change when we consider the BE effects. When considering the first BE corrections, the BE amplification factor reduces to
\begin{equation}\label{FBE}
\fl \qquad F_{\rm BE} \simeq \left[1+\omega^{-2} \left(\frac{\Im(A_{1,{\rm lens}})^2}{2A_{0,{\rm lens}}^2}+\frac{A_{2,{\rm lens}}}{A_{0,{\rm lens}}} \right)\right]\exp\left(\ii \omega^{-1} \frac{\Im(A_{1,{\rm lens}})}{A_{0,{\rm lens}}}\right)\,.
\end{equation}
In order to find the first correction to the phase we need to obtain $A_{1,{\rm lens}}$, whereas to obtain the first correction to the amplitude we must also compute $A_{2,{\rm lens}}$. Given the complexity of the task, we focus hereafter on the determination of $A_{1,{\rm lens}}$, hence on the phase distortion due to BE effects.

\subsection{Newman-Penrose formalism}\label{SecNP}

The Newman-Penrose (NP) formalism involves projecting tensorial quantities onto a null tetrad, allowing one to work exclusively with scalar components. The main formulas and expressions of the NP formalism are collected in appendix B of \cite{Ashtekar:2000hw}, but we collect here the ones that we need on the present article. Starting from the null wave vector, we introduce a null tetrad $\{k^a,n^a,m^a,\bar{m}^a\}$ which is a set of null vectors where the last two are complex-valued. The only non-vanishing scalar products are
\begin{equation}
k_a n^a=-1\,,\qquad m_a\bar{m}^a=1\,,
\end{equation}
and the metric can be expressed as 
\begin{equation}\label{Eq:gab}
g_{ab}=-2k_{(a}n_{b)}+2m_{(a}\bar{m}_{b)}\,.
\end{equation}
With any pair $k^a,n^a$ we define a set of observer velocities indexed by the parameter $w$ as
\begin{equation}\label{Defua}
u_{(w)}^a \equiv \frac{w}{2} k^a + \frac{1}{w} n^a\,,
\end{equation}
which are all related via boosts in the spatial direction of $k^a$. The vectors $m^a,\bar{m}^a$ are orthogonal to these velocities ($m_a u_{(w)}^a=\bar{m}_a u_{(w)}^a=0$) and are thus polarization vectors for all these observers.

The NP scalars are twelve complex spin coefficients formed from the components of the Ricci rotation coefficients (or the tetrad covariant derivative), and their general definitions are 
\begin{eqnarray}
\myfl & \rho \equiv - m^a\bar{m}^b\nabla_bk_a, \hspace{0,3cm} \lambda\equiv\bar{m}^a\bar{m}^b\nabla_bn_a,\hspace{0,3cm} \alpha\equiv-\frac{1}{2}(n^a\bar{m}^b\nabla_bk_a-\bar{m}^a\bar{m}^b\nabla_bm_a), \\
\myfl & \sigma\equiv-m^am^b\nabla_bk_a, \hspace{0,3cm} \mu\equiv\bar{m}^am^b\nabla_bn_a,\hspace{0,3cm} \beta\equiv-\frac{1}{2}(n^am^b\nabla_bk_a-\bar{m}^am^b\nabla_bm_a), \\
\myfl & \tau\equiv-m^an^b\nabla_bk_a,\hspace{0,5cm} \nu\equiv\bar{m}^an^b\nabla_bn_a,\hspace{0,3cm} \gamma\equiv-\frac{1}{2}(n^an^b\nabla_bk_a-\bar{m}^an^b\nabla_bm_a), \\
\myfl & \kappa\equiv-m^ak^b\nabla_bk_a, \hspace{0,5cm} \pi\equiv\bar{m}^ak^b\nabla_bn_a ,\hspace{0,4cm} \epsilon\equiv-\frac{1}{2}(n^ak^b\nabla_bk_a-\bar{m}^ak^b\nabla_bm_a)\,.
\end{eqnarray}
Equivalently the covariant derivative of tetrad vectors is only expressed in terms of the NP scalars, see~\eqref{Eq:nablatetrad1}--\eqref{Eq:nablatetrad4}.

Curvature in vacuum is described by five complex scalars encoding the Weyl projections, defined by
\begin{eqnarray}\label{DefPsin}
 \myfl   & \Psi_0\equiv C_{abcd}k^am^bk^cm^d \,,\quad \Psi_1\equiv C_{abcd}k^an^bk^cm^d \,,\quad \Psi_2\equiv C_{abcd}k^am^b\bar{m}^cn^d \,, \\
 \myfl    & \Psi_3\equiv C_{abcd}k^an^b\bar{m}^cn^d \,,\quad \Psi_4\equiv C_{abcd}n^a\bar{m}^bn^c\bar{m}^d \,,
\end{eqnarray}
where $C_{abcd}$ is the Weyl tensor. When matter is present, we must also consider the components of the Ricci tensor. These ten degrees of freedom are organized as four real valued and three complex scalars, which are
\begin{eqnarray}
\fl\quad \Phi_{00}\equiv\frac{1}{2}R_{ab}k^ak^b \,, \hspace{0,2cm} \Phi_{11}\equiv\frac{1}{4}R_{ab}(k^an^b+m^a\bar{m}^b) \,, \hspace{0,2cm} \Phi_{22}\equiv\frac{1}{2}R_{ab}n^an^b \,, \hspace{0,2cm} \Lambda\equiv\frac{R}{24} \,, \\
\fl \quad \Phi_{01}\equiv\frac{1}{2}R_{ab}k^am^b =\bar{\Phi}_{10} \,, \hspace{0,2cm} \Phi_{02}\equiv\frac{1}{2}R_{ab}m^am^b =\bar{\Phi}_{20} \,, \hspace{0,2cm} \Phi_{12}\equiv\frac{1}{2}R_{ab}m^an^b =\bar{\Phi}_{21}.
\end{eqnarray}
Finally, we define operators which correspond to the projection of covariant derivatives on the null tetrad. The usual notation is
\begin{equation}\label{Defds}
D \equiv k^a\nabla_a\,,\quad \Delta \equiv n^a\nabla_a\,,\quad \delta \equiv m^a\nabla_a\,,\quad \bar{\delta} \equiv \bar{m}^a\nabla_a\,,
\end{equation}
and one must be careful not to confuse $\Delta$ with a Laplacian (in which case we use $\Dlap$), nor $\delta$ with a small variation.
The structure equations for the NP scalars, which are collected in~\cite{Newman:1961qr,Newman:2009}, are obtained by the projection on tetrads of the commutation of two covariant derivative using \eqref{Eq:nablatetrad1}--\eqref{Eq:nablatetrad4}.

\subsection{Choice of null tetrad}\label{SecChoiceTetrad}

The first null vector $k_a$ is chosen to be the eikonal wavevector which is parallel transported, hence 
\begin{equation}\label{Conditionkappa}
\myfl \kappa = \Re(\epsilon)=0 \,,\qquad\Im(\rho)=0\,,\qquad \rho=-\frac{1}{2}\nabla^a k_a\,,\qquad \tau=\bar{\alpha}+\beta\,.
\end{equation}
If $\Im(\epsilon) \neq 0$, we can always redefine the basis as ${\rm e}^{\ii \varphi_m} m^a$ and ${\rm e}^{-\ii \varphi_m} \bar{m}^a$ with $D \varphi_m = - 2\Im(\epsilon)$ such that $\epsilon = 0$. This latter condition corresponds to approximate parallel transport of the polarization vectors, since their variation along the geodesic are then only proportional to $k^a$. It corresponds to a Sachs basis in the context of weak-lensing~\cite{Fleury:2015hgz}. The amount of non-parallel transport is then determined by the value of $\pi$. If $\pi=0$, then the whole tetrad is parallel transported, and not just $k^a$.

In practice, once a geodesic with the associated $k^a$ is chosen to solve for~\eqref{TransportAn}, it can be easy to find one for which $\epsilon=0$, but in general it is not fully parallel transported as we might have $\pi\neq 0$. Let us call $\tilde{k}^a,\tilde{n}^a,\tilde{m}^a,\bar{\tilde{m}}^a$ such tetrad. We then use it to build one which is fully parallel transported (such that $\pi=0$) with the transformation
\begin{equation}\label{Eq:tetradtransfo}
\fl \qquad k^a\equiv\tilde{k}^a \,, \quad m^a \equiv \tilde{m}^a+\alphaG \tilde{k}^a \,, \quad n^a \equiv \tilde{n}^a+\bar{\alphaG} \tilde{m}^a+ \alphaG\bar{\tilde{m}}^a+\alphaG\bar{\alphaG} \tilde{k}^a \,.
\end{equation}
This transformation amounts to changing the class of observers defined by~\eqref{Defua} thanks to a boost in a direction proportional to $\bar{\alphaG} \tilde{m}^a+ \alphaG\bar{\tilde{m}}^a$ which is orthogonal to the spatial direction of $k^a$.\footnote{A velocity vector $u_{(w)}^a$ is obtained via a boost defined by $v^a = (\bar{\alphaG} \tilde{m}^a+ \alphaG\bar{\tilde{m}}^a)/\tilde{w}$, where $\tilde{w}^2 = w^2 + 2 \alphaG \bar{\alphaG}$, applied on $\tilde{u}_{(\tilde{w})}^a$, that is $u_{(w)}^a = (\tilde{u}_{(\tilde{w})}^a + v^a)/\sqrt{1-v_b v^b}$.} The relation between both tetrads is similar to a gauge transformation of the electromagnetic vector field polarization (see e.g. appendix B of \cite{Cusin:2024git}) as the new set of polarization vectors is orthogonal to the new class of observers built according to~\eqref{Defua} with the same $k^a$. The fully parallel transported polarization basis $m^a,\bar{m}^a$ corresponds to the Sachs basis associated with a set of Synge observers~\cite{Cusin:2024git} whose four-velocities are related by parallel transport along the reference geodesic and thus built from~\eqref{Defua} with a constant $w$ along the geodesic.

The spin scalars $-\rho$ and $\sigma$ are the optical scalars (they are called expansion rate and shear rate respectively) of the weak-lensing formalism, and are invariant under such change of tetrad (due to the condition $\kappa=0$). More generally, the shape of an infinitesimal geodesic bundle does not depend on the observer~\cite{Fleury:2015hgz}. The other NP scalars are not invariant under the tetrad transformation~\eqref{Eq:tetradtransfo}, and in particular we find $\pi = \tilde{\pi} + D \bar{\alphaG}$.  The condition of parallel transport, $Dn^a=0$, which is also $\pi =0$, leads to the condition 
\begin{equation}\label{Eq:alpha}
D\alphaG=-\bar{\tilde{\pi}}\,,
\end{equation}
and this is the differential equation that needs to be solved to obtain the $\alphaG$ needed in~\eqref{Eq:tetradtransfo}. 

Finally, the Weyl and Ricci scalars transform as 
\begin{equation}\label{Eq:TransfoPsiPhi}
\fl \qquad {\Psi}_n=\sum_{p=0}^n  {n \choose p} \bar{\alphaG}^p\tilde{\Psi}_{n-p} \,,\quad {\Phi}_{nn'}=\sum_{p=0}^n \sum_{p'=0}^{n'} {n \choose p} {n' \choose p'} \bar{\alphaG}^p\alphaG^{p'}\tilde{\Phi}_{n-p,n'-p'} \,,
\end{equation}
in particular $\Psi_0=\tilde{\Psi}_0$ and $\Phi_{00}=\tilde{\Phi}_{00}$.

\subsection{Geodesic deviation subsystem}

The trajectory of the reference null geodesic $x^a(s)$ satisfies by definition $\dd x^a/\dd \affine = k^a$ hence
\begin{equation}
\frac{\dd}{\dd \affine} = k^a \nabla_a = D \,.
\end{equation}
We aim to express all equations as ordinary differential equations in the operator $D$ so as to integrate all quantities as functions of $\affine$. The evolution of $\rho$ and $\sigma$ is independent of the one of the other NP scalars. They evolve along the reference geodesic according to the structure equation
\begin{equation}\label{DrhoDsigma}
D\rho=\rho^2+\sigma\bar{\sigma}+\Phi_{00} ,\qquad D\sigma= 2\rho\sigma +\Psi_0\,. 
\end{equation}

\begin{figure}[!ht]
\centering
    \includegraphics[width=.85\linewidth]{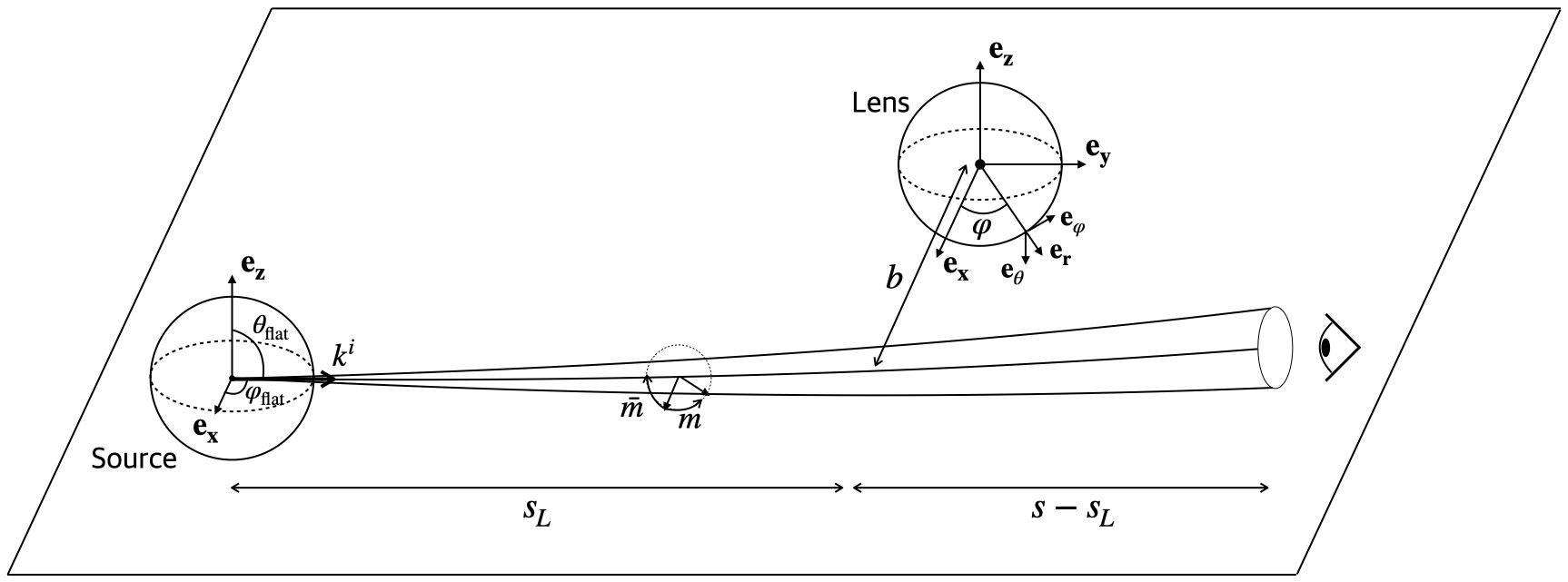}  
\caption{\small The main system of coordinates is a spherical coordinates system centered on the lens in which we express the metric of the lenses [Eqs.~\eqref{BHmetric} and~\eqref{TOVmetric}]. If the geodesic was not deviated, the closest radial distance when $\affine =\affine_{\rm L}$ would be the impact parameter $b$. When placing initial conditions close to the source, we assume the region is very well described by a flat space time and we use a system of spherical coordinates centered on the source, as detailed in~\ref{App:Flat}.}
\label{FigNotation}
\end{figure}

Since close to the source the solution is close to the Minkowski solution (see~\ref{App:Flat}), then $\rho \simeq -1/s$ and $\sigma \simeq 0$. It is therefore impossible to integrate the previous system from the source location at $s=0$. We can either start at a finite $s$ very close to the source, or, as we chose, we can solve for the evolution of the Jacobi matrix $J_{ab}$, which is regular at $s=0$, and then build the solutions for $\rho$ and $\sigma$, see~\ref{App:Jacobi}. The geometry of the geodesic bundle is depicted in Fig.~\ref{FigNotation}.

\subsection{Transport of amplitude and its gradients}

The evolution of $A_0$, given by~\eqref{eqA0}, depends only on $\rho$ as with~\eqref{Conditionkappa} it is also written
\begin{equation}\label{ev0}
DA_0=\rho A_0\,.
\end{equation}
This states that the amplitude in the GO evolves as the inverse of the areal distance $\chi$ defined by $\chi^2 = {\rm det} J_{ab}$ (see~\ref{App:Jacobi}), since this distance satisfies
\begin{equation}\label{Dchi}
D \chi = -\rho \chi\,,\quad \Rightarrow\quad D(A_0 \chi) = 0\,.
\end{equation}
Hence when solving for $A_0$ we can use the results of the Jacobi matrix evolution to deduce $A_0$. Alternatively we can start very close but at finite distance from the source with $\rho = -1/s$, $A_0 = {\rm cst}/s$ and $\sigma=0$, and solve~\eqref{ev0} jointly with~\eqref{DrhoDsigma}.

The theoretical work would stop here if we were interested in weak-lensing, that is in the deformation of a geodesic bundle in the GO approximation. However, we want to integrate the equation~\eqref{TransportAn}, which with definitions~\eqref{Defds} is
 \begin{equation}\label{GeneralDAn}
D\left(\frac{A_n}{A_0}\right)=\frac{\ii}{2}\frac{\square A_{n-1}}{A_0}\,,
\end{equation}
hence we need to know the various gradients of the amplitude. 

We can write the covariant derivative of any scalar $S$ as 
\begin{equation}\label{DecompositionScalaire}
\nabla_a S = -k_a \Delta S - n_a D S+ \bar{m}_a \delta S + m_a \bar{\delta} S \,.
\end{equation}
In order to obtain the transport equation of projected derivatives of $S$ along the reference geodesic, we use for instance
\begin{equation}
D \delta S = \delta D S + [D,\delta] S\,.
\end{equation}
The first term on the r.h.s is expressed with the transport equation of the scalar ($D S = \dots$), and the second term is evaluated from the commutators~\eqref{Eq:commutators}.

For instance, starting from~\eqref{ev0} we get
\begin{eqnarray}\label{eq:gradA}
& D(\Delta A_0) =\rho\Delta A_0 +(\Delta\rho-(\gamma+\bar{\gamma})\rho)A_0 +\bar{\tau}\delta A_0+\tau\bar{\delta}A_0 \,,\label{eq:gradA1} \\
& D(\delta A_0) =2\rho\delta A_0 +(\delta\rho-\tau\rho)A_0 +\sigma\bar{\delta}A_0 \,, \label{eq:gradA2}\\
& D(\bar{\delta} A_0) =2\rho\bar{\delta} A_0+(\bar{\delta}\rho-\bar{\tau}\rho)A_0 +\bar{\sigma}\delta A_0 \,.\label{eq:gradA3}
\end{eqnarray}
The components associated with two derivatives\footnote{Definitions~\eqref{Defds} mean that for instance $\delta \delta S \equiv m^a \nabla_a (m^b \nabla_b S) \neq m^a m^b \nabla_a \nabla_b S$.} are evolving according to
\begin{eqnarray}\label{eq:gradgradA}
 \fl D(\delta \delta A_0) &=&3\rho\delta\delta A_0 + A_0(\bar{\mu}-\mu)\rho\sigma + A_0\rho\tau^2 - 2\bar{\beta}\sigma\delta A_0 - 3\rho\tau\delta A_0 + \sigma\bar{\tau}\delta A_0 - 2A_0\tau\delta\rho \nonumber\\
 \fl &+& 3\delta A_0\delta\rho - A_0\rho\delta\tau + A_0\delta\delta\rho + 2\sigma\delta\bar{\delta}A_0 + 2\beta\sigma\bar{\delta}A_0 - 2\sigma\tau\bar{\delta}A_0 + \delta\sigma\bar{\delta}A_0 \,, \label{eq:gradgradA1}\\
 \fl D(\delta \bar{\delta} A_0) &=&3 \rho \delta \bar{\delta} A_0 + A_0 \rho \tau \bar{\tau} - \bar{\sigma} \tau \delta A_0 - \rho \bar{\tau} \delta A_0 - A_0 \bar{\tau} \delta \rho + \delta A_0\delta\bar{\sigma} - A_0 \rho \delta \bar{\tau} + \bar{\sigma} \delta \delta A_0 \nonumber\\
 \fl &&+ A_0 \delta \bar{\delta} \rho - 2 \rho \tau \bar{\delta} A_0 + 2 \delta \rho \bar{\delta} A_0 - A \tau \bar{\delta} \rho + \delta A_0\bar{\delta} \rho + \sigma \bar{\delta} \bar{\delta} A_0 \,.\label{eq:gradgradA2}
\end{eqnarray}
These components are needed to compute $\square A_0$. Indeed, with~\eqref{Eq:gab} we get 
\begin{eqnarray}
\fl \square A_0&=&g^{ab}\nabla_a\nabla_b A_0 = -2\Delta (DA_0) + 2(\Delta k^a)(\nabla_a A_0) + 2 \delta(\bar{\delta}A_0) -2(\delta\bar{m}^a)(\nabla_a A_0)\nonumber\\
\fl &=& 2(\gamma+\bar{\gamma})\rho A_0 - 2 \mu\rho A_0- 2\Delta\rho A_0 +2\delta\bar{\delta}A_0 +4\beta\bar{\delta}A_0-4\tau\bar{\delta}A_0-2\bar{\tau}\delta A_0 \,,
\end{eqnarray}
where the final expression is obtained with the commutators \eqref{Eq:commutators}, the decomposition of a tetrad covariant derivative~\eqref{Eq:nablatetrad1}--\eqref{Eq:nablatetrad4} and \eqref{DecompositionScalaire}.

\subsection{Source of the first BE correction}

Finally, from~\eqref{GeneralDAn} with $n=1$, the source term for $\Im(A_1)/A_0$ is
\begin{equation} \label{eq:A1}
\myfl D\left(\frac{\Im(A_1)}{A_0}\right)  = (\gamma+\bar{\gamma})\rho - \mu\rho - \Delta\rho + \frac{\delta\bar{\delta}A_0}{A_0} +2(\beta- \tau)\frac{\bar{\delta}A_0}{A_0}- \bar{\tau}\frac{\delta A_0}{A_0} \,.
\end{equation}
This can be recast in a more convenient form in which the operator $\Delta$ does not appear. Using the NP structure equation 
\begin{equation}
\Psi_2=\bar{\delta}\tau-\Delta\rho-\bar{\mu}\rho+(\gamma+\bar{\gamma})\rho+\sigma\lambda+2(\bar{\beta}-\bar{\tau})\tau-2\Lambda
\,,
\end{equation}
this becomes
\begin{eqnarray}\label{DA1Final}
\myfl D\left(\frac{\Im(A_1)}{A_0}\right) &=& 2\Lambda+\Psi_2 - \bar{\delta}\tau+ \frac{\delta\bar{\delta}A_0}{A_0}+(\bar{\mu}-\mu)\rho \nonumber\\
\myfl &+& \sigma\lambda -2(\bar{\beta}-\bar{\tau})\tau  +2(\beta- \tau)\frac{\bar{\delta}A_0}{A_0}- \bar{\tau}\frac{\delta A_0}{A_0}\,.
\end{eqnarray}
We need to know all the scalars and gradients of the source term (the r.h.s of the previous expression) hence we need to solve for their evolution and for the evolution of all expressions which are involved in the process, among which the NP scalars and some of their gradient components.

\subsection{Evolution of NP scalars and their gradients}

The NP structure equations which dictate the evolution of the other NP scalars are
\begin{eqnarray}
\myfl  &&D\mu = \rho\mu +\sigma\lambda +\Psi_2+2\Lambda \,,\qquad D\tau = \tau\rho+\bar{\tau}\sigma +\Psi_1+\Phi_{01}\,,\nonumber\\
\myfl  &&D\lambda = \lambda \rho + \mu \bar{\sigma}+\Phi_{20} \,, \qquad\quad\quad D\beta=\rho\beta+(\bar{\tau}-\bar{\beta})\sigma +\Psi_1\,,\label{eq:spin1}
\end{eqnarray}
and
\begin{equation}
\myfl  D\gamma=\tau\bar{\tau}-\tau\bar{\beta}+\bar{\tau}\beta+\Psi_2+\Phi_{11}-\Lambda \,,\qquad D\nu= \bar{\tau}\mu+\tau\lambda+\Psi_3+\Phi_{21}\,.\label{eq:spin2}
\end{equation}
Close to the source we set initial conditions deduced from the flat space result in the equatorial plane (see~\ref{App:Flat}), for which $\mu \simeq -1/(2s)$ and $\beta=0$.

Applying the same method to obtain the evolution of gradients we get for the one gradient components
\begin{eqnarray}
        \myfl D\delta\rho &=&3\rho\delta\rho-\tau\rho^2 -\sigma\bar{\sigma}\tau +\sigma\bar{\delta}\rho +\bar{\sigma}\delta\sigma+\sigma\delta\bar{\sigma} +\delta\Phi_{00}-\tau\Phi_{00} \,, \label{eq:gradspi1}\\ 
        \myfl D\delta\bar{\tau} &=&2\rho\delta\bar{\tau}+\delta\bar{\Psi}_1 -\rho\tau\bar{\tau}+\bar{\tau}\delta\rho-\tau\bar{\Psi}_1+\sigma\bar{\delta}\bar{\tau}+\bar{\sigma}\delta\tau+\tau\delta\bar{\sigma} -\bar{\sigma}\tau^2 \nonumber\\
        \myfl&&+\delta\bar{\Phi}_{01}-\tau\bar{\Phi}_{01} \,, \label{eq:gradspi2}\\  
        \myfl D\delta\tau &=&2\rho\delta\tau+\delta\Psi_1 -\rho\tau^2+\tau\delta\rho-\tau\Psi_1+\sigma\bar{\delta}\tau+\sigma\delta\bar{\tau}+\bar{\tau}\delta\sigma -\sigma\tau\bar{\tau} \nonumber\\
        \myfl&&+\delta\Phi_{01}-\tau\Phi_{01} \,, \label{eq:gradspi3}\\     
        \myfl D\delta\sigma &=& 3\rho \delta\sigma -\Psi_0 \tau - 2\rho \sigma \tau + \delta\Psi_0 + 2\sigma \delta\rho +  \sigma \bar{\delta}\sigma\,,\label{eq:gradspi4}\\
        \myfl D\delta\bar\sigma &=& 3\rho \delta\bar\sigma-\bar\Psi_0 \tau- 2\rho \bar\sigma \tau+ \delta\bar\Psi_0 + 2\bar\sigma \delta\rho+  \sigma \bar{\delta}\bar\sigma\,.\label{eq:gradspi5}
\end{eqnarray} 

We also need the two gradient components evolutions. Some equations are redundant by complex conjugation. For instance, we can get $\delta\bar{\delta}\bar{\sigma}$ taking the conjugate of $\delta\bar{\delta}\sigma$ and commuting $\delta$ and $\bar{\delta}$ with~\eqref{Commuteddbar}. 
\begin{eqnarray}
\fl        D\delta\bar{\delta}\rho&=& 4\rho \delta\bar{\delta}\rho + \rho^2 \tau \bar{\tau} - \bar{\sigma} \tau \delta \rho - 2\rho \bar{\tau} \delta \rho + \delta\rho\delta\bar{\sigma} - \rho^2 \delta\bar{\tau} + \bar{\sigma} \delta\delta\rho - 3\rho \tau\bar{\delta}\rho + 3\delta\rho\bar{\delta}\rho \nonumber\\
\fl && + \sigma \bar{\delta}\bar{\delta}\rho+\sigma\bar{\sigma}\tau\bar{\tau}-\bar{\sigma}\bar{\tau}\delta\sigma-\sigma\bar{\tau}\delta\bar{\sigma}  -\sigma\bar{\sigma}\delta\bar{\tau}+\bar{\sigma}\delta\bar{\delta}\sigma+\sigma\delta\bar{\delta}\bar{\sigma}-\bar{\sigma}\tau\bar{\delta}\sigma-\sigma\tau\bar{\delta}\bar{\sigma}\nonumber\\
\fl &&+\delta\bar{\sigma}\bar{\delta}\sigma+\delta\sigma\bar{\delta}\bar{\sigma} +\delta\bar{\delta}\Phi_{00}-\Phi_{00}\delta\bar{\tau}-\bar{\tau}\delta\Phi_{00}-\tau\bar{\delta}\Phi_{00}+\tau\bar{\tau}\Phi_{00} \,, \label{eq:gradgradspin1}\\   
\fl         D\delta\delta\rho&=& 4\rho\delta\delta\rho + (\bar{\mu}-\mu)(\rho^2+\sigma\bar{\sigma})\sigma + \rho^2\tau^2 - 2\bar{\beta}\sigma \delta\rho - 5\rho\tau \delta\rho + \sigma\bar{\tau}\delta\rho + 3\delta\rho\delta\rho \nonumber\\
\fl &&- \rho^2 \delta\tau + 2\sigma \delta\bar{\delta}\rho  + 2\beta\sigma \bar{\delta}\rho - 2\sigma\tau \bar{\delta}\rho + \delta\sigma\bar{\delta}\rho +\sigma\bar{\sigma}\tau^2-2\bar{\sigma}\tau\delta\sigma-2\sigma\tau\delta\bar{\sigma}\nonumber\\
\fl && +2\delta\sigma\delta\bar{\sigma}-\sigma\bar{\sigma}\delta\tau+\bar{\sigma}\delta\delta\sigma+\sigma\delta\delta\bar{\sigma} \nonumber \\
\fl && +\delta\delta\Phi_{00}-\Phi_{00}\delta\tau-2\tau\delta\Phi_{00}+\tau^2\Phi_{00}+(\bar{\mu}-\mu)\sigma\Phi_{00} \,,
\end{eqnarray}
\begin{eqnarray}
\fl         D\delta\delta\sigma &=& 4\rho\delta\delta\sigma + (\bar{\mu}-\mu)\sigma\Psi_0+2(\bar{\mu}-\mu)\rho\sigma^2+\Psi_0\tau^2+2\rho\sigma\tau^2-2\tau\delta\Psi_0 - 4\sigma\tau\delta\rho \nonumber\\
\fl &&- 2\bar{\beta}\sigma\delta\sigma - 5\rho\tau\delta\sigma + \sigma\bar{\tau}\delta\sigma + 5\delta\rho\delta\sigma - \Psi_0\delta\tau - 2\rho\sigma\delta\tau + \delta\delta\Psi_0 + 2\sigma\delta\delta\rho \nonumber\\
\fl &&+ 2\sigma\delta\bar{\delta}\sigma + 2\beta\sigma\bar{\delta}\sigma - 2\sigma\tau\bar{\delta}\sigma + \delta\sigma\bar{\delta}\sigma \,, \\
\fl      D\delta\bar{\delta}\sigma&=&
4\rho\delta\bar{\delta}\sigma + \delta\bar{\delta}\Psi_0
+ \delta\sigma\delta\bar{\sigma} + 
2 \delta\bar{\delta}\rho\sigma - 
2\delta\bar{\tau} \rho \sigma - \delta\bar{\tau} \Psi_0
+ \sigma\bar{\delta}\bar{\delta}\sigma + 2\delta\sigma \bar{\delta}\bar{\rho} - 2\sigma\tau \bar{\delta}\bar{\rho}\nonumber\\
\fl &&+ 
3 \delta\rho \bar{\delta}\sigma - 
3 \rho\tau\bar{\delta}\sigma - \tau \bar{\delta} \Psi_0 +\delta\delta\sigma \bar{\sigma} - \delta\sigma\tau\bar{\sigma} - \delta\Psi_0 \bar{\tau} - 2 \delta\sigma\rho\bar{\tau} \nonumber\\
\fl &&- 2 \delta\rho\sigma\bar{\tau} + 2 \rho \sigma \tau\bar{\tau} + \tau\Psi_0\bar{\tau} \,,
\end{eqnarray}
\begin{eqnarray}
  \fl       D\delta\delta\bar{\sigma}&=& 4\rho\delta\delta\bar{\sigma} + (\bar{\mu}-\mu)\bar{\Psi}_0\sigma + 2(\bar{\mu}-\mu)\rho\sigma\bar{\sigma} + \bar{\Psi}_0\tau^2 + 2\rho\bar{\sigma}\tau^2 - 2\tau\delta \bar{\Psi}_0 - 4\bar{\sigma}\tau\delta \rho \nonumber\\
  \fl &&- 2\bar{\beta}\sigma\delta\bar{\sigma}- 5\rho\tau\delta\bar{\sigma} + \sigma\bar{\tau}\delta\bar{\sigma}+ 5\delta\rho\delta\bar{\sigma} - \bar{\Psi}_0\delta\tau - 2\rho\bar{\sigma}\delta\tau + \delta\delta\bar{\Psi}_0 +2\bar{\sigma}\delta\delta\rho\nonumber\\
  \fl && +2\sigma\delta\bar{\delta}\bar{\sigma} +2\beta\sigma\bar{\delta}\bar{\sigma}-2\sigma\tau\bar{\delta}\bar{\sigma}+\delta\sigma\bar{\delta}\bar{\sigma} \,, \\
\fl         D\delta\bar{\delta}\bar{\sigma}&=&
4\rho\delta\bar{\delta}\bar{\sigma} + \delta\bar{\delta}\bar{\Psi}_0
+ \delta\bar{\sigma}\delta\bar{\sigma} +\sigma\bar{\delta}\bar{\delta}\bar{\sigma} + 2\delta\bar{\sigma} \bar{\delta}\bar{\rho} + 
3 \delta\rho \bar{\delta}\bar{\sigma} - 
3 \rho\tau\bar{\delta}\bar{\sigma} - \tau \bar{\delta}\bar{\Psi}_0 + 2 \delta\bar{\delta}\rho\bar{\sigma} \nonumber\\
\fl && +\bar{\sigma}\delta\delta\bar{\sigma} - 
2\delta\bar{\tau} \rho\bar{\sigma} - \delta\bar{\sigma}\tau\bar{\sigma} - 2\bar{\sigma}\tau \bar{\delta}\bar{\rho} - \delta\bar{\Psi}_0 \bar{\tau} - 2 \delta\bar{\sigma}\rho\bar{\tau} \nonumber\\
\fl &&- 2 \delta\rho\bar{\sigma}\bar{\tau} + 2 \rho \bar{\sigma} \tau\bar{\tau} - \delta\bar{\tau}\bar{\Psi}_0 + \tau\bar{\tau}\bar{\Psi}_0 \,.\label{eq:gradgradspin6}
\end{eqnarray} 
If the geodesic bundle remains in vacuum, then the Ricci components vanish, and the Weyl components only source directly the shear and its gradient components, whereas the expansion rate and its gradient components are indirectly sourced by the former ones.

\subsection{Weyl and Ricci components}

The explicit form of the Weyl and Ricci tensors are obtained once the metric is known, hence it depends on the specific case in which we apply the previous formalism. If we use a non-fully parallel transported tetrad, that is the tilded tetrad of section~\eqref{SecChoiceTetrad}, we would obtain the $\tilde{\Psi}_n$ and $\tilde{\Phi}_{nm}$ components, and we then only need to use~\eqref{Eq:TransfoPsiPhi} to get the $\Psi_n$ and $\Phi_{nm}$ associated with the fully parallel transported tetrad. For this, we need to integrate \eqref{Eq:alpha} along the geodesic so as to know the $\alphaG$ involved in the tetrad transformation. 

Some of the equations also include the projected derivatives of the Weyl and Ricci scalars. Let us see on an example how we can handle these more complicated term systematically. 
From the definition of $\delta\Psi_1$, we get 
\begin{eqnarray}
\fl \delta\Psi_1&=&m^p\nabla_p(C_{abcd}k^an^bk^cm^d) =m^p\nabla_p(C_{abcd})k^an^bk^cm^d+C_{abcd}m^p\nabla_p(k^an^bk^cm^d) \cr
\fl & = &m^p\nabla_p(C_{abcd})k^an^bk^cm^d +\mu\Psi_0+2\beta\Psi_1-3\sigma\Psi_2 \,,
\end{eqnarray}
where to obtain the last line we used~\eqref{Eq:nablatetrad1}--\eqref{Eq:nablatetrad4}. Hence when the NP and Weyl scalars are already known, we need only explicit expressions of the the tetrads components of $\nabla_p C_{abcd}$, which can always be found once the metric and the null tetrad are known. Again, in practice we would obtain these expressions in the non-fully parallel transported tetrad (the tilded tetrad), but knowing the transformation property of the tetrads~\eqref{Eq:tetradtransfo}, we can find the components in the parallel transported tetrad. The other expressions involving covariant derivatives of the Weyl or Ricci tensor which are needed to compute $A_1$ are collected in~\ref{App:Weyl}.

\subsection{Summary of BE differential system}\label{SecSummary}

Let us summarize the equations which we need to solve along the reference geodesic.
\begin{enumerate}
\item We first solve for the trajectory of the reference geodesic as a function of $\affine$ and determine the tilded tetrad and its associated $\tilde{\pi}$.
\item We then solve~\eqref{Eq:alpha} to determine $\alphaG$ needed to build the fully parallel transported tetrad.
\item We solve the subsystem~\eqref{DrhoDsigma} along with~\eqref{Dchi} to obtain $\rho$, $\sigma$ and $A_0$. Alternatively we can deduce them after solving the Jacobi matrix equation with~\eqref{EqJacobiComponent}.
\item We solve for the evolution of $\lambda$, $\mu$, $\tau$ and $\beta$ dictated by Eqs.~\eqref{eq:spin1}. Since the scalar $\nu$ and $\gamma$ are not needed to obtain $A_1$ we do not need to solve Eqs.~\eqref{eq:spin2}.
\item We then solve for the gradients of the needed scalars along $m^a$ and $\bar{m}^a$, that is we solve~\eqref{eq:gradspi1}--\eqref{eq:gradspi5} and ~\eqref{eq:gradA2}--\eqref{eq:gradA3} 
\item We also solve for the gradients of these gradient components, that is~\eqref{eq:gradgradspin1}-\eqref{eq:gradgradspin6} and \eqref{eq:gradgradA1}--\eqref{eq:gradgradA2}.
\item Eventually this allows to obtain the sources in~\eqref{DA1Final} and solve it to obtain $A_1/A_0$.
\end{enumerate}
In practice, all coupled ordinary differential equations can be integrated at the same time for simplicity and the non-vanishing initial conditions that we set close to the source are obtained from the flat space solutions of~\ref{App:Flat} in the equatorial plane.

\subsection{Weak-field approximation}

The previous program is considerably reduced if we use the weak-field approximation, that is reducing to first order in $G$. The weak-field approximation is interesting for two reasons. First it provides the leading effect. Second this allows to refine the initial conditions close to the source by including the first corrections beyond the Minkowski solution of~\ref{App:Flat}. In the weak-field approximation the metric is described by
\begin{equation}\label{Weakgab}
\dd s^2=-\left(1+2\phi\right)\dd t^2+\left(1-2\phi\right)\delta_{ij}\dd x^i\dd x^j \,.
\end{equation}
The non-vanishing components of the Ricci and Weyl tensors are~\cite{Kling:2008ph} at linear order 
\begin{eqnarray}\label{RCfirst}
\fl &&R_{00}\simeq\Dlap\phi \,,\quad R_{ij} \simeq \delta_{ij}\Dlap\phi \,,\nonumber\\
&&C_{i0j0} \simeq \partial_i\partial_j\phi-\frac{1}{3}\delta_{ij}\Dlap\phi \,,\quad C_{ijkl} \simeq 4\delta_{[i[p}\partial_{j]}\partial_{q]}\phi \,.
\end{eqnarray}
As curvature tensors vanish at lowest order, we only need to use the flat space tetrad of~\ref{App:Flat} in order to evaluate their components.  We find
\begin{equation}\label{Psi012thin}
 \Psi_0^{(1)} = 2 \delta \delta \phi\,,\quad \Psi_1^{(1)} = - \partial_s \delta \phi\,,\quad \Psi_2^{(1)} = \frac{1}{2}\partial_{ss}\phi - \frac{1}{6}\Dlap \phi\,,
\end{equation}
\begin{equation}\label{Phi012thin}
\Phi_{00}^{(1)} = 4 \Phi_{11}^{(1)} = 12 \Lambda^{(1)} = \Dlap \phi\,,\quad \Phi_{10}^{(1)}=\Phi_{20}^{(1)}=\Phi_{12}^{(1)}=0\,.
\end{equation}
Applying definitions~\eqref{DefPsin} and~\eqref{Defds} with~\eqref{FlatSpherical}, we also find
\begin{equation}
\bar{\delta} \Psi_1^{(1)}  = \frac{3}{2 s}\partial_{ss}\phi + \frac{1}{2}\partial_{sss}\phi -\frac{1}{2s}\Dlap \phi -\frac{1}{2}\partial_s \Dlap \phi\,.
\end{equation}
At first order we also get the useful identity
\begin{equation}\label{Deltaddbar}
\Dlap \phi = (\delta \bar{\delta} + \delta \bar{\delta})\phi+ \partial_{ss}\phi+\frac{2}{s}\partial_s \phi\,.
\end{equation}
It is very useful because in vacuum the Ricci tensor vanishes, implying $\Dlap\phi=0$.

First, it can be seen that in the weak-field approximation $\mu = \bar{\mu}$. Furthermore, the only non-vanishing quantities for a spherical source in flat spacetime (when $\phi=0$) are $\rho=-1/\affine$, $\mu$ and $A_0={\rm cst}/\affine$, hence when solving at first order we only need to solve with these background functional forms the reduced system
\begin{eqnarray}
&&D\delta\bar{\tau}=2\rho\delta\bar{\tau}+\delta\bar{\Psi}_1^{(1)}+ {\cal O}(G^2)\,,\label{Simplified1}\\
&&D\delta\bar{\delta}\rho=4\rho\delta\bar{\delta}\rho-\rho^2\delta\bar{\tau}+\delta\bar{\delta}\Phi_{00}^{(1)} + {\cal O}(G^2)\,,\\
&&D\delta\bar{\delta}A_0=3\rho\delta\bar{\delta}A_0+A_0\delta\bar{\delta}\rho-\rho A_0\delta\bar{\tau}+ {\cal O}(G^2)\,,\label{Simplified3}\\
&& D\left(\frac{\Im(A_1)}{A_0}\right) = \frac{\delta\bar{\delta}A_0}{A_0}- \bar{\delta}\tau +2\Lambda^{(1)}+\Psi_2^{(1)} +{\cal O}(G^2)\,,
\end{eqnarray}
where we used the short notation ${\cal O}(G^2) \equiv {\cal O}(\phi^2)$. 

\section{Weyl lensing}\label{SecWeyl}

We now apply the previous program for a geodesic bundle which crosses only vacuum, hence the Ricci components vanish identically. We consider two useful approximations in this context. First we restrict the equations and solutions to the weak-field regime, that is at order $G$, and we shall show that in vacuum there is no BE effect at that order. We also consider the case of a thin lens for which the non-vanishing Weyl components are confined to a small region around the lens, which is much smaller than the distances between the lens and the source or the distance between the lens and the observer. We obtain the first non-vanishing correction at order $G^2$. In order to check the accuracy of these approximations we solve the BE set of equations for a geodesic crossing the outer region of a static black hole.

\subsection{Weak-field approximation in vacuum}

In vacuum, which we assume hereafter, $\Dlap\phi=0$ hence in particular we get
\begin{equation}\label{Psi2weak}
\left.\Psi_2^{(1)} \right|_{\rm vac}=\frac{1}{2}\partial_{ss}\phi \,,\quad \left.\delta\bar{\Psi}_1^{(1)}\right|_{\rm vac} =\frac{3}{2\affine}\partial_{ss}\phi+\frac{1}{2}\partial_{sss}\phi \,.
\end{equation}
The general solutions at order $G$ to the system~\eqref{Simplified1}--\eqref{Simplified3} with $\delta \bar{\delta} \Phi_{00}=0$ are 
\begin{equation}\label{AnalyticIC}
\myfl \left.\delta\bar{\tau}\right|_{G}=\frac{\mathcal{I}}{\affine^2}\,,\quad \left.\delta\bar{\delta}\rho\right|_{G}=-\frac{1}{\affine^4}\int_0^\affine \mathcal{I}(\affine_1)\dd\affine_1\,,\quad\left.\delta\bar{\delta}A_0\right|_{G}=\frac{A_0}{\affine^3}\int_0^\affine \mathcal{I}(\affine_1)\dd\affine_1\,,
\end{equation}
where
\begin{eqnarray}
\myfl&&\mathcal{I}\equiv\left.\int_0^\affine \affine_1^2\delta\bar{\Psi}_1^{(1)}(\affine_1)\dd\affine_1\right|_{\rm vac} = \frac{1}{2}\affine^2\partial_{ss}\phi+\frac{1}{2}\affine\partial_{s}\phi-\frac{1}{2}(\phi-\phi_{\rm ini})\,,\\
\myfl&&\frac{1}{\affine}\int_0^\affine \mathcal{I}(\affine_1)\dd\affine_1=\frac{1}{2}\affine \partial_{s}\phi-\frac{1}{2}(\phi-\phi_{\rm ini})\,.
\end{eqnarray}

\subsection{Vanishing of BE corrections at order $G$}\label{SecTheorem}

At first order in the potential the evolution of $\Im(A_1)$ is dictated by the first line of~\eqref{DA1Final} with $\Lambda=0$ and $\mu=\bar{\mu}$, that is 
\begin{equation}
D\frac{\Im(A_1)}{A_0}=\left.\frac{\delta\bar{\delta}A_0}{A_0}\right|_{G}-\left.\delta\bar{\tau}\right|_{G}+\Psi_2^{(1)}+ {\cal O}(G^2)\,.
\end{equation}
Using the previous solutions~\eqref{AnalyticIC} and~\eqref{Psi2weak}, and applying a few integrations by parts, we obtain
\begin{eqnarray}\label{Cancellation}
\myfl \frac{\Im(A_1)}{A_0}&=&\int_0^\affine\frac{1}{\affine_1^2} \left( \frac{1}{\affine_1}\int_0^{\affine_1} \mathcal{I}(\affine_2)\dd\affine_2 - \mathcal{I}(\affine_1)  \right)\dd\affine_1 + \int_0^\affine\Psi_2(\affine_1)\dd\affine_1 \nonumber+ {\cal O}(G^2)\\
\fl &=& {\cal O}(G^2)\,.
\end{eqnarray}
Therefore in vacuum there is no contribution at order $G$ in $\Im(A_1)/A_0$. Furthermore, since $\square A_n$ is the source of $A_{n+1}$, there will be no order-$G$ contribution to any $A_n$ with $n \geq 1$. Given these precise cancellations, the numerical resolution is also improved if we use the first order approximations~\eqref{AnalyticIC} to fix initial conditions close to the source. This result is in sharp constrast with Eq.~(51) of \cite{CarrilloGonzalez:2025gqm} where a frequency-dependent phase shift $\propto GM/(\omega b^2)$ at the lens level is obtained from a correspondence with partial wave expansions.

\subsection{Thin-lens approximation}\label{ThinLens}

Assuming that the support of the Weyl scalars $\Psi_n$ is very localized with respect to the scales we consider, we can perform the thin-lens approximation to get approximate analytical solution which are not restricted to the first order in the potential. We only consider the solution for $\affine > \affine_{\rm L}$ where the latter affine parameter is the one at which the infinitely thin less is reached from the source. In addition, $\Psi_0$ is generically the dominant Weyl scalar since, as seen in~\eqref{Psi012thin}, it does not involve any derivative along the affine parameter which vanishes after integration. In the thin-lens description, it reduces to
\begin{equation}
\Psi_0\simeq \mathcal{I}_{\Psi_0}\delta(\affine-\affine_{\rm L}) \quad \text{with}\quad \mathcal{I}_{\Psi_0}=\int^{+\infty}_{-\infty}\Psi_0(s) \dd s \,.
\end{equation}
We will also assume that the axis of polarization have been chosen such that $\Psi_0$, and thus $\mathcal{I}_{\Psi_0}$, are real.
The weak lensing subsystem~\eqref{DrhoDsigma} has known solutions 
\begin{eqnarray}\label{Solrho}
\rho &=& -\frac{(1-\mathcal{I}_{\Psi_0}\affine_{\rm L})}{2[\affine-\affine_{\rm L} \mathcal{I}_{\Psi_0}(\affine-\affine_{\rm L})]}-\frac{(1+\mathcal{I}_{\Psi_0}\affine_{\rm L})}{2[\affine+\affine_{\rm L} \mathcal{I}_{\Psi_0}(\affine-\affine_{\rm L})]}  \,,\\
\sigma &=& \frac{(1+\mathcal{I}_{\Psi_0}\affine_{\rm L})}{2[\affine+\affine_{\rm L} \mathcal{I}_{\Psi_0}(\affine-\affine_{\rm L})]}-\frac{(1-\mathcal{I}_{\Psi_0}\affine_{\rm L})}{2[\affine-\affine_{\rm L} \mathcal{I}_{\Psi_0}(\affine-\affine_{\rm L})]}\,,
\end{eqnarray}
and the previous choice leads to $\sigma=\bar{\sigma}$. 

We can then solve the equation of the other variables in the thin-lens approximation using the results for $\rho$ and $\sigma$ and the variation of constants. First, to solve for $\mu$ and $\lambda$, we define the new variables $X\equiv\lambda-\mu$ and $Y\equiv\lambda+\mu$ which verify the equation $DX=(\rho-\sigma)X$ and $DY=(\rho+\sigma)Y$, and we find
\begin{eqnarray}
\mu=-\frac{1}{4[\affine+\mathcal{I}_{\Psi_0}\affine_{\rm L}(\affine-\affine_{\rm L})]}-\frac{1}{4[\affine-\mathcal{I}_{\Psi_0}\affine_{\rm L}(\affine-\affine_{\rm L})]} \,,\\
\lambda=\frac{1}{4[\affine+\mathcal{I}_{\Psi_0}\affine_{\rm L}(\affine-\affine_{\rm L})]}-\frac{1}{4[\affine-\mathcal{I}_{\Psi_0}\affine_{\rm L}(\affine-\affine_{\rm L})]}\,.
\end{eqnarray}

We then wish to iterate the process to obtain analytic expressions of gradient components in the thin-lens approximation. We first assume for simplicity that 
\begin{equation}
\delta \Psi_0 \simeq \mathcal{I}_{\delta\Psi_0} \delta(\affine-\affine_{\rm L}) \quad \text{with}\quad \mathcal{I}_{\delta\Psi_0}=\int^{+\infty}_{-\infty}\delta\Psi_0(s) \dd s\,,
\end{equation}
is real, which implies $\delta\sigma=\bar{\delta}\bar{\sigma}$. Furthermore at first order in the potential, $\delta \bar{\Psi}_0 \propto \delta \bar{\delta} \bar{\delta} \phi$. And with the first order relation~\eqref{Deltaddbar}, we get that the integral of $\delta \bar{\Psi}_0$ is subdominant, so we neglect it. This implies in particular that $\delta \bar{\sigma} = \bar{\delta} \sigma$. Note however that $\delta\sigma\neq\delta\bar{\sigma}$ hence we need to solve for $\delta\rho$, $\delta\sigma$ and $\delta\bar{\sigma}$ whose evolution is given by the close set of equations~\eqref{eq:gradspi1}, \eqref{eq:gradspi4} and~\eqref{eq:gradspi5}. Defining the variables $X\equiv 2\delta\rho+\delta\sigma+\delta\bar{\sigma}$, $Y\equiv 2\delta\rho-\delta\sigma-\delta\bar{\sigma}$ and $Z\equiv \delta\rho+\delta\sigma-2\delta\bar{\sigma}$, we get the equations $DX=3(\rho+\sigma)X+\delta\Psi_0$, $DY=(3\rho-\sigma)Y-\delta\Psi_0$ and $DZ=(3\rho-\sigma)Z+\delta\Psi_0$. After solving for $X,Y,Z$, we deduce $\delta\rho$, $\delta\sigma$ and $\delta\bar{\sigma}$. For instance,
\begin{eqnarray}\label{drho}
\delta\rho&=&\frac{\affine\affine_{\rm L}^4(\affine-\affine_{\rm L})\mathcal{I}_{\Psi_0}\mathcal{I}_{\delta\Psi_0}}{[\affine-\affine_{\rm L} \mathcal{I}_{\Psi_0}(\affine-\affine_{\rm L})]^3[\affine+\affine_{\rm L} \mathcal{I}_{\Psi_0}(\affine-\affine_{\rm L})]^2} \,.
\end{eqnarray} 
Iterating this process, it is possible to obtain analytic expressions for the sources of $\Im(A_1)/A_0$ in~\eqref{DA1Final}. However expressions become exceedingly huge, hence we do not report their explicit forms.

It is clear on~\eqref{Solrho} that a caustic ($\rho \to \infty$) arises when $|\mathcal{I}_{\Psi_0}| \affine_{\rm L} \geq 1$. It takes place when $|\mathcal{I}_{\Psi_0}|L =1$ where we defined the length
\begin{equation}
\myL \equiv \frac{\affine_{\rm L}(\affine-\affine_{\rm L})}{\affine} \quad \Leftrightarrow \quad \frac{1}{L} = \frac{1}{\affine_{\rm L}} + \frac{1}{\affine-\affine_{\rm L}}\,.
\end{equation}
When a caustic arises, our formalism cannot be applied since it also leads to a divergence of gradients, e.g. $\delta \rho$ in~\eqref{drho}, in contradiction with what we assumed initially. Therefore this motivates us to consider a weak-lensing regime in which $|\mathcal{I}_{\Psi_0}| \affine_{\rm L} <1$ so as to perform an expansion in $G$. As we have already proven that $A_1$ vanishes identically if we restrict to the first order, we need to keep at least contributions quadratic in $G$.

\subsection{First BE correction at order $G^2$}\label{SecFirstG2}

The dominant source at order $G^2$ for $\delta\bar{\delta}\rho$ is the quadratic term $\delta\sigma \bar\delta\bar\sigma$. The first order evolution of $\delta \sigma$ is dictated by $D \delta \sigma = 3 \rho \delta \sigma + \delta \Psi_0$ whose solution is
\begin{equation}
\delta\sigma \simeq \mathcal{I}_{\delta\Psi_0} \left(\frac{\affine_{\rm L}}{\affine}\right)^3+{\cal O}(G^2)\,.
\end{equation}
We then get 
\begin{eqnarray}\label{ddrhoddAG2}
\myfl &&\delta\bar{\delta}\rho\simeq \left.\delta\bar{\delta}\rho\right|_{G}+|\mathcal{I}_{\delta\Psi_0}|^2 \left(\frac{\affine_{\rm L}}{\affine}\right)^5(\affine-\affine_{\rm L})\,,\nonumber\\
\myfl &&\frac{\delta\bar{\delta}A_0}{A_0}\simeq \left.\frac{\delta\bar{\delta}A_0}{A_0}\right|_{G}+\frac{1}{2}|\mathcal{I}_{\delta\Psi_0}|^2\left(\frac{\affine_{\rm L}}{\affine}\right)^4(\affine-\affine_{\rm L})^2\,.
\end{eqnarray}
The last term is main contribution to the sources of $A_1$, hence we obtain
\begin{equation}\label{CentralResult1}
\myfl \quad\boxed{ \frac{\Im(A_1)}{A_0}\simeq \frac{1}{6}|\mathcal{I}_{\delta\Psi_0}|^2 L^3 + {\cal O}(G^3)= \frac{1}{6}|\mathcal{I}_{\delta\Psi_0}|^2\left(\frac{\affine_{\rm L}}{\affine}\right)^3(\affine-\affine_{\rm L})^3 + {\cal O}(G^3)\,.}
\end{equation}
This is the first central result of this article.

\subsection{Schwarzschild lensing}

\subsubsection{Reference null geodesic}

We want to check the previous approximations with a numerical solution, hence we consider lensing around a static black hole. We describe the background metric of the point lens by the Schwarzschild metric
\begin{equation}\label{BHmetric}
\fl\hspace{1,5cm} \dd s^2=-\left(1-\frac{R_{\rm S}}{r}\right)\dd t^2+\left(1-\frac{R_{\rm S}}{r}\right)^{-1}\dd r^2+r^2(\dd\theta^2+\sin^2\theta \dd\varphi^2) \,,
\end{equation}
where the Schwarzschild radius is $R_{\rm S} \equiv 2GM$. The associated Killing vector fields are $\xi^a_1=(\partial_t)^a$ and $\xi^a_2=(\partial_\varphi)^a$, and the quantities $\xi_{1,2}^a k_a$ are constant along a given null geodesic defined by $k^a$. So the conserved quantities along a geodesic are the impact parameter \begin{equation}\label{Defb}
k_a\xi^a_1=r^2 \frac{\dd \varphi}{\dd \affine}=b\,,
\end{equation}
and energy $k_a\xi^a_2=-(1-R_{\rm S}/r)\dd t /\dd \affine=-1$. Note that with our convention $k^a$ is dimensionless and the physical wave vector is $k^a_{\rm phys} = \omega k^a$. Restricting for simplicity to the equatorial plane ($\theta=\pi/2$) the null condition reads 
\begin{equation}
\fl \qquad g_{ab}k^a k^b=-(1-R_{\rm S}/r)\left(\frac{\dd t}{\dd \affine}\right)^2+(1-R_{\rm S}/r)^{-1}\left(\frac{\dd r}{\dd \affine}\right)^2+r^2\left(\frac{\dd \varphi}{\dd \affine}\right)^2=0\,.
\end{equation}
Replacing the conserved quantities, we get 
\begin{equation}\label{Eqdrds}
\frac{\dd r}{\dd \affine}= \mp \sqrt{1-\frac{b^2}{r^3}(r-R_{\rm S})} \,,
\end{equation}
where here and in subsequent expressions the upper sign (resp. lower sign) is for the first part (resp. second part) of the plane. In practice~\eqref{Eqdrds} and~\eqref{Defb} are better solved by translating them as equations for $\dd x/\dd \affine$ and $\dd y/\dd \affine$ from which $r,\varphi$ are deduced, starting from the point of closest approach which is in general given by
\begin{equation}\label{rminexact}
r_{\rm min} = \frac{2b}{\sqrt{3}}\cos\left[\frac{1}{3}\arccos\left(-\frac{3^{3/2} R_{\rm S} }{2 b} \right) \right] .
\end{equation}
The result of numerical integrations is also checked against the approximations of~\ref{App:SZO1}.

\subsubsection{Choice of null tetrad}

Let us introduce a null tetrad for this metric. We define 
\begin{equation}
{\bm e}_x=\frac{b}{r}f(r) \partial_r\pm\frac{U(r)}{r} \partial_\phi\,,\quad {\bm e}_y=\frac{1}{r} \partial_\theta\,.
\end{equation}
where $f(r) \equiv 1- R_{\rm S}/r$ and $U(r)\equiv \sqrt{1-(b/r)^2f(r)}$, so as to define the polarization vector $\tilde{\bm m}^a \equiv \frac{1}{\sqrt{2}}({\bm e}_x^a+\ii {\bm e}_y^a)$. The other null vectors are
\begin{equation}
 \myfl \tilde{\bm k}= f^{-1}(r)\partial_t\mp U(r)\partial_r + \frac{b}{r^2}\partial_\phi\,,\quad \tilde{\bm n}=\frac{1}{2}\left( \partial_t\pm U(r)f(r)\partial_r -b\frac{f(r)}{r^2}\partial_\phi \right).
\end{equation}
The needed Weyl scalars in this tilded tetrad are
\begin{eqnarray}
    & \tilde\Psi_0= - \frac{3b^2 R_{\rm S}}{ 2r^5} \,,\\
    & \tilde\Psi_1= \mp\frac{3b R_{\rm S}\ }{2\sqrt{2}r^4}\sqrt{1-\frac{b^2}{r^3}(r-R_{\rm S})}\,, \\
    & \tilde\Psi_2= -\frac{R_{\rm S}}{4r^6}\left[2 r^2-3b^2 (r-R_{\rm S})\right] \,.
    \end{eqnarray}
The Weyl components in the parallel transported basis are then deduced with~\eqref{Eq:TransfoPsiPhi}. The equation~\eqref{Eq:alpha}, which determines the tetrad transformation, is
\begin{equation}
D\alphaG=\frac{R_{\rm S} b}{2\sqrt{2}r^3}\,.
\end{equation}
At first order in $G$ we find for this geometry 
\begin{equation}\label{IPsiBH}
\myfl\mathcal{I}_{\Psi_0}= -\frac{2 R_{\rm S}}{b^2}+{\cal O}\Big((GM)^2\Big)\,,\qquad \mathcal{I}_{\delta\Psi_0} =\frac{4 \sqrt{2} R_{\rm S}}{b^3} +{\cal O}\Big((GM)^2\Big)\,.
\end{equation}

Finally let us mention that in a static metric we can relate the derivative $\Delta$ to the other ones. To do so, we write the $\Delta$ derivative in the non parallel transported tetrad basis, where we know the relation between $\tilde{\Delta}$ and $\tilde{D}$, and transform back the derivatives to the parallel tetrad basis. This way we get
\begin{equation}
\Delta=-\left(\alphaG\bar{\alphaG}+\frac{1}{2}f(r)\right)D + \bar{\alphaG}\delta + \alphaG\bar{\delta} \,.
\end{equation}
Thus, the four different projected for a scalars are not independent in a static metric. As we choose to work with~\eqref{DA1Final} and not with~\eqref{eq:A1}, no operator $\Delta$ is involved and we do not need to use the previous relation. However, this relation allowed us to check that by solving \eqref{eq:A1}, where there is a $\Delta \rho$ contribution, we obtain the same result.

\subsection{Numerical comparison}

\begin{figure}[!ht]
\centering
\begin{minipage}{.48\linewidth}
    \centering
    \includegraphics[width=1\linewidth]{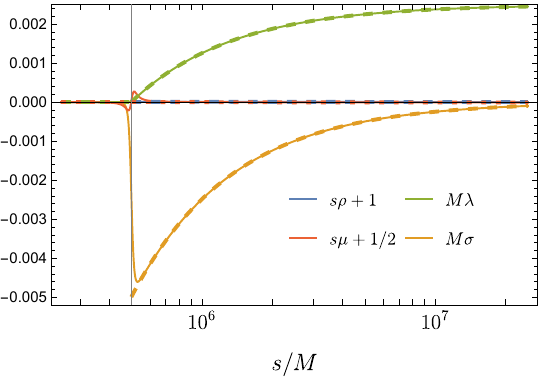}
    \caption{\small Numerical integration (continuous lines) and analytic approximation (dashed lines) for the first NP scalars (parameters given in main text).}
    \label{fig:first}
\end{minipage}\hfill
\begin{minipage}{.48\linewidth}
    \centering
    \includegraphics[width=1\linewidth]{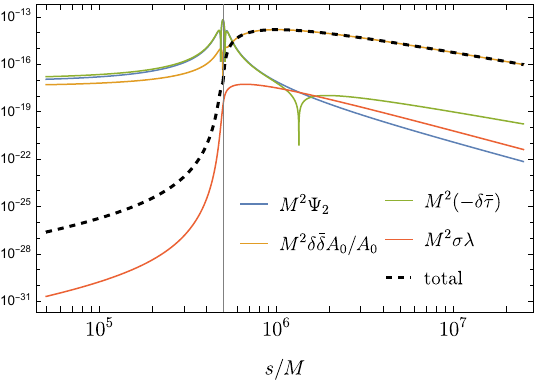}
    \caption{\small The r.h.s. of \eqref{DA1Final} is decomposed into its main contributions (parameters given in main text). The vertical bar is located  at the position of the black hole lens.}
    \label{fig:second}
\end{minipage}
\end{figure}

We now apply our formalism to the case of a static black hole described in the previous section. The free parameters are $M$, $b$ and the initial distance $r_{\rm i}$ of the source which is approximately $\affine_{\rm L}$. Using~\eqref{IPsiBH} in~\eqref{CentralResult1} we obtain the analytic approximation of the lowest BE correction
\begin{equation}\label{FinalA1BH}
\frac{\Im(A_1)}{A_0} \simeq\frac{4}{3}\left(\frac{2 R_{\rm S}}{b^3}\right)^2 L^3=\frac{4}{3}\left(\frac{2 R_{\rm S}}{b^3}\right)^2\left(\frac{\affine_{\rm L}}{\affine}\right)^3(\affine-\affine_{\rm L})^3\,,
\end{equation}
that we wish to compare with the full numerical result. Again at the level of the lens, that is when $\affine = \affine_{\rm L}$, there is no BE correction, that is no extra phase shift of order $G^2/\omega$. This differs from Eq.~(51) of \cite{CarrilloGonzalez:2025gqm} where a frequency dependent phase shift $\propto (GM)^2/(\omega b^3)$ at the lens level is obtained.

\begin{figure}[!ht]
\centering
    \includegraphics[width=0.48\textwidth]{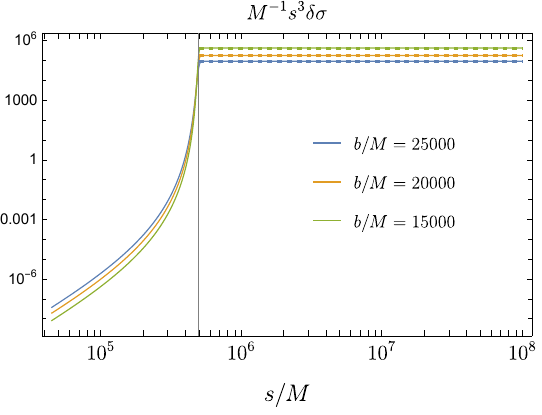}  
    \includegraphics[width=.5\textwidth]{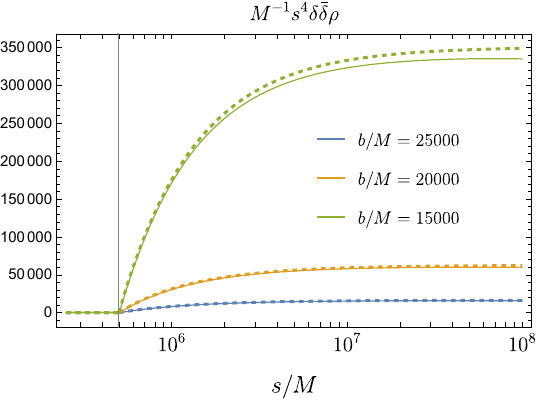}  \\
    \includegraphics[width=.47\textwidth]{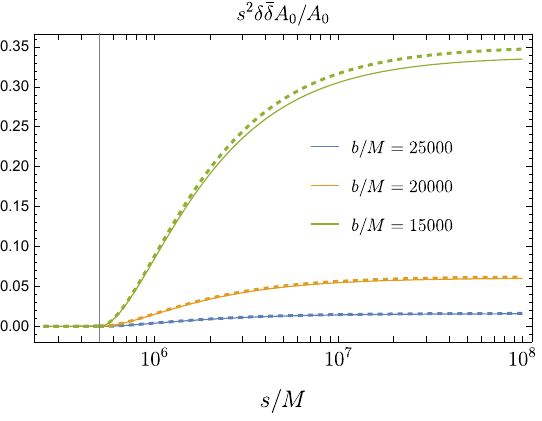}  
    \includegraphics[width=.51\textwidth]{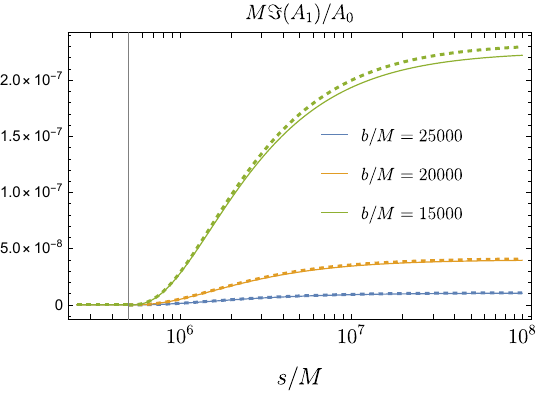}  
\caption{\small The intermediate quantities, needed to estimate the source of $A_1$, are determined numerically (continuous lines) or analytically approximated (dashed lines) for $r_{\rm i}/R_{\rm S}=2.5\times 10^5$ with expressions of section~\eqref{SecFirstG2}.}
\label{weyl}
\end{figure}

Since our formalism only applies in the regime of weak lensing, we ensure that the condition $|\mathcal{I}_{\Psi_0}| L < 1$ is always satisfied. The results for the first NP scalars are plotted in Fig.~\ref{fig:first}. We have set $b/R_{\rm S}=10^4$ and $r_{\rm i}/R_{\rm S}=2.5\times 10^5$, hence we are in the weak lensing regime ($|\mathcal{I}_{\Psi_0}|\affine_{\rm L} \simeq 2 (R_{\rm S}/b)^2 (r_{\rm i}/R_{\rm S}) \simeq 5\times 10^{-3} \ll 1$). The main contribution of the r.h.s in~\eqref{DA1Final} are plotted in Fig.~\ref{fig:second}, and it can be seen that the total is much smaller than the individual contributions before and around the lens, due to the precise and exact cancellation [Eq.~\eqref{Cancellation}] at first order in $G$. Nonetheless, after the lens $A_1$ is sourced by terms of order $G^2$ which enhance the dominant source term $\delta \bar{\delta} A_0/A_0$ [Eq.~\eqref{ddrhoddAG2}].

The most important quantities involved in the method are plotted in Fig.~\ref{weyl} for several values of $b/R_{\rm S}$ but the same $r_{\rm i}/R_{\rm S}$. The analytic approximations of section~\ref{SecFirstG2} are very good since we are deep in the weak-lensing regime and also in the linear regime ($|\phi| < GM/b = 5\times10^{-5}$). We also check on the lower right panel that the approximation~\eqref{FinalA1BH} to estimate the lowest BE correction is accurate.

Let us evaluate the size of the first BE correction. We consider a source located at a distance $a R_{\rm S}$ from the lens of mass $M=10^6{\rm M}_\odot$. Even though our derivation has been obtained for a scalar field, let us consider gravitational waves produced by a binary black hole orbiting a supermassive black hole which acts as the lens, and we choose $a=10^3$~\cite{Graham:2020gwr}. For a wavelength typical of LISA ($\lambda = 10^{11}\,{\rm m}$), 
\begin{equation}
\myfl \frac{\Im(A_1)}{\omega A_0} \approx 3\times10^{-2}\times \left(\frac{10^{-1}}{\epsilon}\right)^6 \left(\frac{10^{3}}{a}\right)^3 \left(\frac{\lambda}{10^{11}\,{\rm m}}\right) \left(\frac{10^6{\rm M}_\odot}{M}\right)\,,
\end{equation}
where we wrote the impact parameter as $b=\epsilon\, aR_{\rm S} $, and for a source approximately behind the lens we used $\affine_{\rm L} \simeq aR_{\rm S}$. Note that with this parameterization $\mathcal{I}_{\Psi_0} \affine_{\rm L} = 2/(\epsilon^2 a) = 0.2\,(10^{-1}/\epsilon)^2(10^3/a)$, hence the parameters chosen ensure that the weak-lensing assumption is satisfied.

\section{Ricci lensing}\label{SecRicci}

The cancellation of sources for $A_1$ at order $G$ crucially relies on the vanishing of the Ricci tensor. Hence we now consider a reference geodesic which crosses a non-vanishing matter density.

\subsection{Solutions at order $G$}\label{SecOG1}
 
Let us first consider a weak-field non-vacuum solution with the metric~\eqref{Weakgab} which now satisfies at first order the Poisson equation 
\begin{equation}
\Dlap\phi = 4\pi G\rho_{\rm mat}\,,
\end{equation}
where $\rho_{\rm mat}$ is the matter density field. In section~\ref{SecTheorem}, we demonstrated that, at order $G$, the terms without $\Dlap\phi$ do not contribute to source $\Im(A_1)$. These terms will therefore cancel each other out, and only the $\rho_{\rm mat}$ terms will introduce new contributions. Proceeding as before to solve the system~\eqref{Simplified1}--\eqref{Simplified3}, but this time with a non-vanishing $\delta \bar \delta \Phi_{00} = 4 \pi G\delta \bar \delta \rho_{\rm mat}$, we obtain
\begin{eqnarray}\label{AnalyticIC2}
\fl \quad \left.\delta\bar{\tau}\right|_{G}=\frac{\mathcal{I}_{\rm mat}}{\affine^2}\,,\quad \left.\delta\bar{\delta}\rho\right|_{G}=-\frac{1}{\affine^4}\int_0^\affine \mathcal{I}_{\rm mat}(\affine_1)\dd\affine_1 +\frac{4\pi G}{s^4}\int_0^\affine\affine_1^4\delta\bar{\delta}\rho_{\rm mat}(\affine_1)\dd\affine_1 \,, \\
\fl \quad \left.\delta\bar{\delta}A_0\right|_{G}=\frac{A_0}{\affine^2}\left( \frac{1}{\affine}\int_0^\affine \mathcal{I}_{\rm mat}(\affine_1)\dd\affine_1 +\frac{4\pi G}{s}\int_0^\affine \affine_1^3\left(\affine-\affine_1\right)\delta\bar{\delta}\rho_{\rm mat}(\affine_1)\dd\affine_1 \right) ,\nonumber
\end{eqnarray}
where
\begin{eqnarray}
\myfl&&\mathcal{I}_{\rm mat}\equiv \int_0^\affine \affine_1^2\delta\bar{\Psi}_1^{(1)}(\affine_1)\dd\affine_1=\mathcal{I}-2\pi G\affine^2\rho_{\rm mat} + 2\pi G\int_0^\affine\affine_1\rho_{\rm mat}(\affine_1)\dd\affine_1 \,,\\
\myfl &&\int_0^\affine \mathcal{I}_{\rm mat}(\affine_1)\dd\affine_1 = \int_0^\affine \mathcal{I}(\affine_1)\dd\affine_1+2\pi G\int_0^\affine \affine_1(\affine - 2 \affine_1)\rho_{\rm mat}(\affine_1)\dd\affine_1\,.
\end{eqnarray}
We are left with
\begin{equation}
D\frac{\Im(A_1)}{A_0}=\left.\frac{\delta\bar{\delta}A_0}{A_0}\right|_{G}-\left.\delta\bar{\tau}\right|_{G}+\Psi_2^{(1)}+2\Lambda^{(1)} + {\cal O}(G^2) \,.
\end{equation}
The terms without $\rho_{\rm mat}$ vanish, and $\rho_{\rm mat}$ does not contribute in $\Psi_2^{(1)}+2\Lambda^{(1)}$ thanks to \eqref{Psi012thin} and \eqref{Phi012thin}. Using~\eqref{AnalyticIC2} and integrating by parts, we are left with 
\begin{equation}
\fl\quad \frac{\Im(A_1)}{A_0} = \frac{2\pi G}{\affine^2}\int_0^{\affine}\affine_1^2\rho_{\rm mat}(\affine_1)\dd\affine_1 + \frac{2\pi G}{\affine^2}\int_0^{\affine}\affine_1^2(\affine -\affine_1)^2\delta\bar{\delta}\rho_{\rm mat}(\affine_1)\dd\affine_1 + {\cal O}(G^2)\,.
\end{equation}
Compared to the vacuum case, there is a contribution at order $G$ due to non-vanishing Ricci scalars.
Moreover, if we use the thin-lens approximation to write $\rho_{\rm mat}\simeq\mathcal{I}_{\rho_{\rm mat}} \delta(\affine-\affine_{\rm L})$ and $\delta\bar{\delta}\rho_{\rm mat}\simeq \mathcal{I}_{\delta\bar{\delta}\rho_{\rm mat}} \delta(\affine-\affine_{\rm L})$, we finally get
\begin{equation}\label{FinalBeautiful}
\boxed{\frac{\Im(A_1)}{A_0}\simeq2\pi G\left(\frac{\affine_{\rm L}}{\affine}\right)^2 \Big( \mathcal{I}_{\rho_{\rm mat}} +\mathcal{I}_{\delta\bar{\delta}\rho_{\rm mat}}(\affine-\affine_{\rm L})^2 \Big) + {\cal O}(G^2)\,.}
\end{equation}
This is the second central result of this article. Note that the first term modifies the phase immediately after the thin lens (when $\affine=\affine_{\rm L}$) and decreases afterwards. 

\subsection{Simplified TOV geometry}

In order to check the previous approximate results, we will consider a toy model where a spherically symmetric star is described by a TOV solution with a prescribed matter density profile. In addition we consider for simplicity a beam that is crossing at the center of the transparent star.

\subsubsection{Structure of the star}

The TOV metric is in general
\begin{equation}\label{TOVmetric}
\myfl \dd s^2=-e^{2\phi(r)}\dd t^2+\left(1-\frac{2 G m(r)}{r}\right)^{-1}\dd r^2+r^2(\dd\theta^2+\sin^2\theta \dd\varphi^2) \,.
\end{equation}

The Tolman–Oppenheimer–Volkoff equations, obtained from this metric via Einstein equations and the hydrodynamic equilibrium, govern the structure of the star, and are
\begin{equation}\label{TOVeq}
\myfl \frac{{\rm{d}} p}{{\rm{d}} r}=-(p+\rho_{\rm mat})\frac{\rm{d}\phi}{{\rm{d}}r} \,\,\,\,\,,\,\,\,\, \frac{{\rm{d}}\phi}{{\rm{d}}r}=G\frac{m+4\pi r^3 p}{r(r-2Gm)} \,\,\,\,\,\text{and}\,\,\,\, \frac{{\rm{d}}m}{{\rm{d}}r}=4\pi\rho_{\rm mat} r^2 \,,
\end{equation}
where $\rho_{\rm mat}(r)$ is the matter density, $p(r)$ is the matter pressure, $M$ is the total mass of the star and $R$ its radius (such that $m(R)=M$).

\subsubsection{Radial geodesic}

We consider a geodesic following a straight line with $b=0$ and passing through $r=0$ inside this "transparent" star. The null geodesic condition leads to 
\begin{equation}
\frac{\dd r}{\dd t}=\mp e^{\phi}\sqrt{f(r)}\,,\quad \Rightarrow\quad \frac{\dd r}{\dd\affine}=\mp e^{-\phi}\sqrt{f(r)} \,,
\end{equation}
where $f(r)\equiv 1-2 G m(r)/r$. We introduce the parallel transported null tetrad 
\begin{equation}\label{NulltetradTOV}
\fl  {\bm k}= e^{-2\phi}\partial_t \mp \sqrt{f(r)}e^{-\phi}\partial_r\,, \quad {\bm n}=\frac{1}{2} \left(\partial_t \pm \sqrt{f(r)}e^{\phi}\partial_r \right)\,, \quad{\bm m}=\frac{1}{\sqrt{2}r}\left( \partial_\theta +\ii\partial_\phi \right)\,.
\end{equation}

\subsection{Reduced BE system}

The symmetries imply that only the spin-$0$ quantities do not vanish, hence several spin NP scalars, namely $\sigma$, $\tau$, $\lambda$ and $\beta$, vanish identically. The reduced system of equations which needs to be solved to follow the program of section~\ref{SecSummary} is
\begin{eqnarray}
D A_0 &=& \rho A_0\,,\label{SysTOV1}\\
D \rho &=& \rho^2 + \Phi_{00}\,,\label{SysTOV2}\\
D \mu &=& \rho \mu + \Psi_2 + 2 \Lambda
\end{eqnarray}
\begin{eqnarray}
&&D \delta \bar{\tau} = 2 \rho \delta \bar{\tau}+ \delta \bar{\Psi}_1 +\delta\bar{\Phi}_{01} \,,\\
&&D \delta \bar\delta \rho = 4 \rho \delta \bar\delta \rho - \rho^2 \delta \bar{\tau} + \delta \bar{\delta} \Phi_{00} - \Phi_{00} \delta \bar{\tau}\,,\label{Dddbrho}\\
&&D \delta \bar \delta A_0 = 3 \rho \delta \bar \delta A_0 - A_0 \rho \delta \bar{\tau} + A_0 \delta \bar \delta \rho\,,\\
&&D\frac{\Im(A_1)}{A_0} = \frac{\delta \bar \delta A_0 }{A_0} - \delta \bar{\tau} + \Psi_2+2\Lambda\,.\label{SysTOVn}
\end{eqnarray}
The only surviving Weyl and Ricci scalars are $\Psi_2$, $\Phi_{00}$, $\Phi_{11}$, $\Phi_{22}$ and $\Lambda$.
Defining
\begin{equation}
H(r) \equiv r^2(\phi'+4\pi r G \rho_{\rm mat})- G m(1+2 r \phi')\,, 
\end{equation}
which reduces to $4\pi G \rho_{\rm mat}\simeq \Dlap \phi$ in weak-field regime, the explicit form of those scalars and the needed derivatives are obtained by considering the general expressions of \ref{App:Weyl} with the particular metric~\eqref{TOVmetric} and null tetrad~\eqref{NulltetradTOV}, and are
\begin{eqnarray}
\myfl &&\Psi_2 = -\frac{G m}{r^3}+ \frac{4}{3}\pi G\rho_{\rm mat} \,, \\
\myfl &&\delta \bar{\Psi}_1 = -3 \rho \Psi_2\mp\frac{3}{r}{\rm e}^{-\phi}\sqrt{f(r)}\left(-\frac{G m}{r^3}+ \frac{4}{3}\pi G\rho_{\rm mat} \right)\,,\\
\myfl &&\Phi_{00} =4\Phi_{22}=\frac{e^{-2\phi} H}{r^3}\,,\quad \Phi_{11} = \frac{H}{4r^3}\,,\quad \Lambda = -\frac{H}{4r^3}+\frac{4}{3}\pi G\rho_{\rm mat}  \,,\\
\myfl &&D \Phi_{00}=\mp \frac{e^{-3\phi}\sqrt{f(r)}}{r^3}(4\pi r^3 \rho'_{\rm mat}-3 H \phi')\,,\\
\myfl &&\Delta \Phi_{00}=2(\gamma+\bar{\gamma})\Phi_{00}\mp \frac{{\rm e}^{-\phi}\sqrt{f(r)}}{2r^3}(-4\pi r^3 \rho'_{\rm mat}- H \phi')\,,\\
\myfl &&\delta \bar{\delta} \Phi_{00}=\mu D \Phi_{00}- \rho \Delta \Phi_{00} + 2 \Phi_{00}(\gamma+\bar{\gamma}+\bar{\mu})\rho-4 \rho^2 \Phi_{11} \nonumber\\
\myfl&&\qquad -2 \rho (\delta \bar{\Phi}_{01}+\bar{\delta}\Phi_{01}) + f(r){\rm e}^{-2\phi}(4\pi r^3 \rho'_{\rm mat}- H \phi')\,,\\
\myfl &&\delta \bar{\Phi}_{01}= \mu \Phi_{00} -2 \rho \Phi_{11}\,.
\end{eqnarray}
Outside the star, that is for $r>R$, only the Weyl scalar is non-vanishing.

\subsection{Numerical comparison}

We will consider two simple density profiles for which we will compare the numerical implementation of the previous system to the  analytic approximations at order $G$ obtained in section~\ref{SecOG1}.\\

{\it Uniform density profile : }we set $\rho_{\rm mat}(r)=\rho_0$ and $m(r)=4\pi\rho_0r^3/3$ for which there are analytic solutions to~\eqref{TOVeq} which are
\begin{equation}
p(r)=\rho_0\frac{g(r)-1}{3-g(r)} \,,\quad \phi(r)=\ln\left( \frac{1}{2}\left[3-g(r)\right]\sqrt{1-R_{\rm S}/R} \right) ,
\end{equation}
where $g(r)\equiv\sqrt{(1-R_{\rm S}r^2/R^3)/(1-R_{\rm S}/R)}$ and $R_{\rm S} \equiv 2 G M$.
The relation between $\rho_0$ and $M$ is simply given by $M=\frac{4}{3}\pi\rho_0R^3$, and we find 
\begin{equation}\label{IIdd1}
\mathcal{I}_{\rho_{\rm mat}}=2\rho_0R \,,\quad \mathcal{I}_{\delta\bar{\delta}\rho_{\rm mat}}=0 \,,
\end{equation}
which must be used in~\eqref{FinalBeautiful} to assess the first BE correction. Inside the star, $\Psi_2$ is exactly zero. However, outside the star it is the only non-vanishing Weyl or Ricci NP scalar.\\

{\it Parabolic density profile : } we set $\rho_{\rm mat}(r)=\rho_0(1-c\,r^2/R^2)$, where $c$ is a free constant parameter. We obtain $m(r)=\frac{4}{3}\pi\rho_0 r^3(1-\frac{3}{5}c\,r^2/R^2)$ and so $M=\frac{4}{15}(5-3c)\pi\rho_0 R^3$. The potential and pressure are solved numerically from the TOV equations~\eqref{TOVeq}. In this case we must use in~\eqref{FinalBeautiful}
\begin{equation}\label{IIdd2}
\mathcal{I}_{\rho_{\rm mat}}=2\left(1 - \frac{c}{3}\right)\rho_0 R  \,,\quad \mathcal{I}_{\delta\bar{\delta}\rho_{\rm mat}}=-4c \frac{\rho_0}{R} \,.
\end{equation}
Contrary to the uniform density case, $\Psi_2$ does not vanish inside the star. \\

In the thin lens approximation we can solve~\eqref{SysTOV2} similarly to section \ref{ThinLens}. After the lens we obtain
\begin{equation}\label{SolutionRhoThinLens}
\rho=-\frac{(1-\bar{\kappa})}{[s-\bar{\kappa}(s-s_{\rm L})]}\,,\qquad \bar{\kappa}\equiv4\pi G{\cal I}_{\rho_{\rm mat}}s_{\rm L}\,.
\end{equation}
Therefore, we ensure that $\bar{\kappa}<1$ to avoid caustics and remain in the weak lensing regime.

\begin{figure}[ht]
    \includegraphics[width=.49\linewidth]{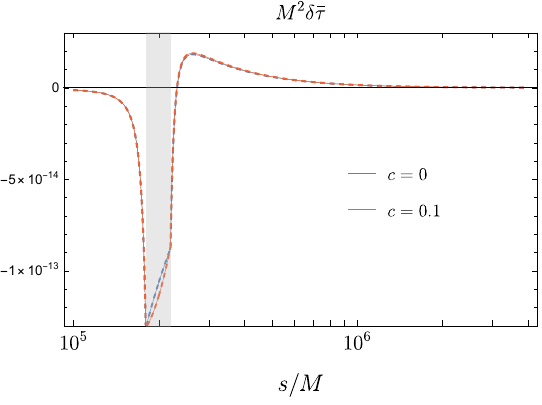}  
    \includegraphics[width=.49\linewidth]{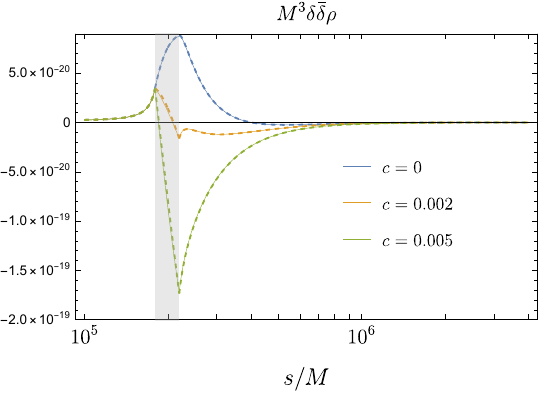} \\ 
    \includegraphics[width=.49\linewidth]{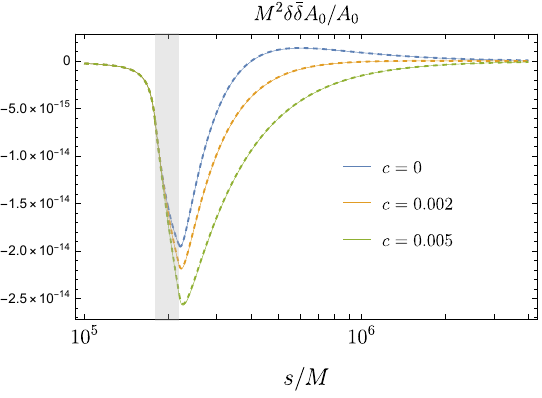}  
    \includegraphics[width=.49\linewidth]{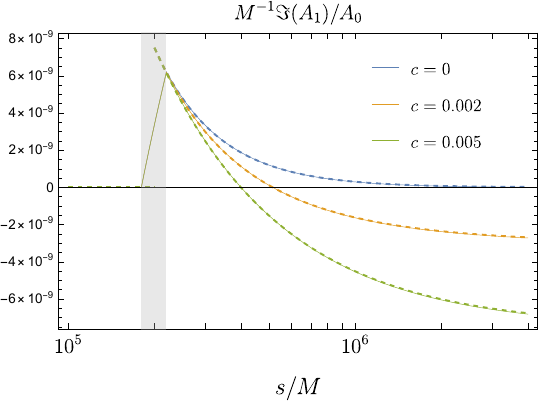}  
\caption{\small Intermediate quantities and $\Im(A_1)/A_0$ are solved either numerically (continuous lines) or analytically approximated (dashed lines) with expressions reported in section~\ref{SecOG1} for $R/R_{\rm S}=10^4$ and $r_{\rm i}/R_{\rm S}=2.5\times10^5$. The vertical bands denote the star region with $\rho_{\rm mat} \neq 0$.}
\label{figTOV}
\end{figure}

In Fig.~\ref{figTOV} we compare the analytic approximations of section~\ref{SecOG1} with the full numerical results obtained with the system~\eqref{SysTOV1}--\eqref{SysTOVn}, and it can be checked that there is a very good agreement since we have chosen a regime for which $\phi \ll 1$. When the matter density is constant, then from~\eqref{IIdd1} the second contribution in~\eqref{FinalBeautiful} vanishes. In that case the first contribution tends to vanish when $\affine \gg \affine_{\rm L} $. However, whenever the reference geodesic experiences a non-vanishing two-dimensional Laplacian (in the plane orthogonal to the direction of propagation) of the matter density, that is if $\delta \bar\delta \rho_{\rm mat} \neq 0 $, the second contribution does not vanish as seen in~\eqref{IIdd2} for the parabolic profile example. Therefore $\Im(A_1)/A_0$ settles to a constant value far from the lens. This second contribution is typically much larger than the first one, unless there is a strict uniform matter density.  

Let us evaluate the size of the first BE correction for a wave passing through a small galaxy that we model with a parabolic density with $c=1$. Considering a wavelength typical of LISA ($\lambda = 10^{11}\,{\rm m}$), an observer far from the lens ($\affine \gg \affine_{\rm L}$), a source at $100\,{\rm Mpc}$ behind the lens of mass $10^9\,{\rm M}_\odot$ and radius $R=1\,{\rm kpc}$, we obtain
\begin{equation}
\myfl -\frac{\Im(A_1)}{\omega A_0} \approx 4\times10^{-6}\times \left(\frac{M}{10^{9} {\rm M}_{\odot}}\right) \left(\frac{1\,{\rm kpc}}{R}\right)^4 \left(\frac{s_{\rm L}}{100\,{\rm Mpc}}\right)^2 \left(\frac{\lambda}{10^{11}{\rm m}}\right)\,.
\end{equation}
Note that these parameters correspond to the weak-lensing regime since $\bar{\kappa}=5\times 10^{-2}$.

\section{Discussion and conclusion}\label{SecConlusion}

The BE expansion allows us to obtain corrections to the geometric optics approximation without relying on the diffraction integral. We have computed the first correction in power of $1/\omega$, which amounts to a modification of the phase, as seen on Eq.~\eqref{FBE}, which cannot be absorbed as a frequency independent time delay, and we found two non-trivial results. To compare them with previous literature let us first introduce the notation~\cite{Jow:2022pux}
\begin{equation}\label{Defkappaomega}
\bar \kappa \equiv \Sigma L\,,\qquad \bar \omega \equiv \omega b^2/L\,,
\end{equation}
where $\Sigma \equiv 4\pi G {\cal I}_{\rho_{\rm mat}}$ for the Ricci case, in agreement with definition~\eqref{SolutionRhoThinLens}, and $\Sigma \equiv 4 G M /b^2$ for lensing around a black hole. The length $b$ is the characteristic scale of the matter density gradients in the Ricci case, and the impact parameter for lensing around a black hole. 

First when the reference geodesic only travels through vacuum (Weyl lensing), there is no BE effect at order $GM/\omega$ and the first non-vanishing BE effect starts at order $(GM)^2/\omega$. With the notation~\eqref{Defkappaomega}, $F_{\rm BE}=A^{\rm lens}/A^{\rm lens}_0$ obtained in Eq.~\eqref{FinalA1BH} is of order $\bar \kappa^2/\bar \omega$. Note that our result differs from the second line of Eq.~(51) in \cite{CarrilloGonzalez:2025gqm} since we find that there is no phase shift $\propto GM/\omega$ and that the phase shift at order $(GM)^2/\omega$, given by Eq.~\eqref{FinalA1BH}, vanishes at the lens level (when $\affine = \affine_{\rm L}$). The partial wave expansion relies on a heuristic identification $b \omega = \ell+1/2$, whereas with the NP formalism we are able to track exactly the dynamics of the first BE correction, and thus able to pinpoint the precise cancellations which lead to the vanishing of the $GM/\omega$ correction. 

Second, when the reference geodesic travels through a non-vanishing matter density (Ricci lensing), we find that a phase shift is generated just after the thin lens (when $\affine = \affine_{\rm L}$). It is proportional to $\Sigma/\omega$ but this effects fades out as $(\affine_{\rm L}/\affine)^2$. 
However there is a second and larger effect in~\eqref{FinalBeautiful} of order $\bar \kappa/\bar \omega$ which vanishes at the lens but does not vanish far from the lens. 

We can also guess the order of magnitude of the subsequent corrections. Given the structure of the recursive construction~\eqref{TransportAn}, corrections from $\omega^{-n}A_n/A_0$ with $n>1$ will bring corrections with additional powers of $1/\bar\omega$. Indeed the operator $D$ on the l.h.s is of order $1/L$ and $\square A_{n-1}/A_0$ on the r.h.s is of order $A_{n-1}/(A_0 b^{2})$ hence $A_n/\omega$ is of order $A_{n-1} L/(b^2 \omega) = A_{n-1}/\bar\omega$. Corrections in vacuum involve a sum of terms of order $\bar\kappa^2/\bar\omega^n$ when restricting to the lowest order in $G^2$, whereas for lensing by gradients of matter density it would be a sum of terms of order $\bar\kappa/\bar\omega^n$ when restricting to corrections of order $G$. We thus infer that for small $\bar\kappa$ a necessary condition for the geometric optics to be valid is $\bar \omega \gg 1$.
The conditions stated in literature on that topic vary and we summarize here how they are related. 

A first condition stated~\cite{Takahashi:2004mc} is $\lambda \equiv 2\pi/\omega \ll R_{\rm S}$ for lensing around a point mass. If the impact parameter is such that we are in the strong lensing regime, then it is of the order of the Einstein impact parameter $b_E \equiv \sqrt{ 2 R_{\rm S} L}$. Since $\bar \omega = 2 R_{\rm S} \omega (b/b_E)^2$ this matches the condition $\bar \omega \gg 1$. However this latter condition authorizes wavelength much larger than $R_{\rm S}$ when $b \gg b_E$. A second condition stated in~\cite{Dalang:2021qhu} is $\lambda \ll b$ with effects on the polarization of gravitational waves of order $R_{\rm S} \lambda/b^2$. This order of magnitude is exactly the one of the first correction in~\eqref{FinalBeautiful} with the choice of $\Sigma$ mentioned above in the case of a black hole. We find however that the amplitude of a scalar wave is also corrected by the second term of~\eqref{FinalBeautiful}, which is much larger and is of order $\bar{\kappa}/\bar \omega$ and for which the condition $\lambda \ll b$ is too permissive. The condition $\bar \omega \gg 1$, also stated as
\begin{equation}
\lambda \ll \frac{b^2}{L}\,,
\end{equation}
is more restrictive. It can also be interpret as the condition that the Fresnel scale $\sqrt{\lambda L}$ is much smaller than the typical lens size set by $b$.

In Ref.~\cite{Jow:2022pux} the condition stated for the validity of the geometric optics is different from the usually accepted condition~\cite{Pen:2013njl,Dong:2018lmm,Er:2021shs} $\bar \omega \gg 1$. It is argued with exact evaluations of the diffraction integral for a particular lens profile, that the condition should rather be $\bar \omega \bar \kappa \gg 1$ for all values of $\bar \kappa$. The potential considered has a non-vanishing Laplacian hence it is to be compared with our results of Ricci lensing. Since our results have been obtained in the weak lensing regime in which $\bar \kappa \ll 1$ by assumption, we do get a different condition. We can nonetheless guess the size of BE corrections if not in the weak lensing regime since our results must be related, with differences though, to the ones obtained with the diffraction integral. When the magnification is not negligible, the length $L$ is replaced by $L \sqrt{\mu_{\rm GO} }$ where $\mu_{\rm GO}  = (s/\chi)^2 = 1/{\rm det}{\cal A}$ is the geometric optics magnification defined in~\ref{App:Jacobi}. Therefore we must replace $\bar \kappa \to \bar \kappa \sqrt{\mu_{\rm GO} }$ and $\bar \omega \to \bar \omega /\sqrt{\mu_{\rm GO} }$ in all our estimates. Notably we expect higher orders to be enhanced by powers of $\sqrt{\mu_{\rm GO} }/\bar \omega$, hence we conjecture that the general condition should be 
\begin{equation}\label{NewCondition}
\bar \omega/\sqrt{\mu_{\rm GO} } \gg 1 \,\quad \Rightarrow \quad \bar\omega (1- \bar \kappa) \gg 1\,,
\end{equation}
where in the second condition we have assumed that there is no shear nor rotation to express the magnification as $\mu_{\rm GO}  = 1/(1-\bar \kappa)^2$. With the condition~\eqref{NewCondition}, the condition on $\bar \kappa \bar\omega$ of Ref.~\cite{Jow:2022pux} becomes $\bar \kappa \bar\omega \gg |\bar \kappa/(1-\bar \kappa)|$. For very strong lensing ($\bar \kappa \gg 1$) this reduces to the condition $\bar \kappa \bar\omega  \gg 1$. However, for weak lensing this is $\bar \omega \gg 1$ which is more permissive than $\bar \kappa \bar\omega  \gg 1$, a trend which can be noted in the left part of Fig. 3 of Ref.~\cite{Jow:2022pux}.

Our method can in principle be extended to vector fields, that is for light, or tensor fields, that is for the propagation of gravitational waves, and it has already been shown that GW polarization is altered by BE effects~\cite{Cusin:2019rmt,Dalang:2021qhu}. We nonetheless expect the derivation of the full differential system for the first BE corrections to be theoretically challenging. In addition, our perturbative approach is restricted to the computation of corrections to the geometric optics regime, and it cannot handle the deep wave-optics regime ($\bar \omega \ll 1$) for which a scattering perspective~\cite{Pijnenburg:2022pug,Pijnenburg:2024btj,Braga:2024pik,CarrilloGonzalez:2025gqm} seems more promising. However the usual diffraction integral, with a frequency-independent gravitational time delay, cannot be a universal starting point to assess the wave optics effects since we have shown that a frequency-dependent phase modulation takes place right after a thin lens.

\section*{Acknowledgements}

We thank Giulia Cusin for motivating this research and for numerous technical discussions on the topic.
We also thank Guillaume Faye, Pierre Fleury, Julien Larena and Jean-Philippe Uzan for discussions and their careful reading of the draft.
\appendix

\section{Null tetrad covariant derivative components}\label{App:NP}

The covariant derivative of the null tetrad is fully express in terms of the NP scalars. With the conditions~\eqref{Conditionkappa}, this reduces to 
\begin{eqnarray} 
\fl \nabla_a k_b &=& -(\gamma+\bar{\gamma})k_a k_b -\bar{\sigma}m_a m_b -\sigma\bar{m}_a\bar{m}_b -2\rho m_{(a}\bar{m}_{b)} +2\tau k_{(a}\bar{m}_{b)} +2\bar{\tau}k_{(a} m_{b)}\nonumber\,,\label{Eq:nablatetrad1}\\
\fl \nabla_a n_b &=& \lambda m_a m_b+\bar{\lambda}\bar{m}_a\bar{m}_b +\mu\bar{m}_a m_b+\bar{\mu}m_a\bar{m}_b -\tau\bar{m}_a n_b -\bar{\tau}m_a n_b \label{Eq:nablatetrad2}\\
\fl&&-\nu k_a m_b -\bar{\nu}k_a\bar{m}_b + (\gamma+\bar{\gamma})k_a n_b \,,\\
\fl \nabla_a m_b &=& (\bar{\tau} -2\bar{\beta})m_a m_b -\bar{\nu}k_a k_b +(2\beta-\tau)\bar{m}_a m_b +\bar{\mu}m_a k_b -\rho m_a n_b \nonumber\\
\fl&&-\sigma\bar{m}_a n_b +\bar{\lambda}\bar{m}_a k_b -(\gamma-\bar{\gamma})k_a m_b+ \tau k_a n_b \,,\\
\fl \nabla_a \bar{m}_b &=& (\tau -2\beta)\bar{m}_a \bar{m}_b -\nu k_a k_b +(2\bar{\beta}-\bar{\tau}) m_a \bar{m}_b +\lambda m_a k_b - \bar{\sigma}m_a n_b \nonumber\\
\fl&&- \rho \bar{m}_a n_b +\mu \bar{m}_a k_b +(\gamma-\bar{\gamma})k_a \bar{m}_b + \bar{\tau} k_a n_b \,.\label{Eq:nablatetrad4}
\end{eqnarray}

Commutators of derivatives along the tetrad vectors are then deduced by repeated application of the previous rules. For instance $[D,\delta]^a = [k,m]^a = k^b \nabla_b m^a - m^b\nabla_b k^a$. We eventually obtain
\begin{eqnarray}\label{Eq:commutators}
& \Delta D-D\Delta =(\gamma+\bar{\gamma})D-\bar{\tau}\delta-\tau\bar{\delta} \,,\\
& \delta D-D\delta =\tau D-\rho\delta-\sigma\bar{\delta}\,, \\
& \delta\Delta-\Delta\delta =-\bar{\nu}D+(\mu-\gamma+\bar{\gamma})\delta+\bar{\lambda}\bar{\delta}\,, \\
& \bar{\delta}\delta-\delta\bar{\delta} =(\bar{\mu}-\mu)D+(\bar{\tau}-2\bar{\beta})\delta-(\tau-2\beta)\bar{\delta}\,.\label{Commuteddbar}
\end{eqnarray}

\section{Weyl components needed in the BE system}\label{App:Weyl}

The components of derivatives of curvature tensors are decomposed as
\begin{eqnarray}
\fl D\Psi_0 = k^a k^b k^p m^c m^d \nabla_b C_{pcad} \\
\fl \Delta\Psi_0 = k^a k^p m^b m^c n^d \nabla_d C_{pbac} +4\Psi_0\gamma - 4\Psi_1\tau \\
\fl \delta\Psi_0= k^a k^p m^b m^c m^d \nabla_d
C_{p b a c} + 4\Psi_0 \beta - 4\Psi_1\sigma \\
\fl \bar{\delta}\Psi_0= k^a k^p m^b m^c \bar{m}^d \nabla_d
C_{p b a c} -4 \Psi_0 \bar{\beta}- 4 \Psi_1 \rho + 
 4 \Psi_0\bar{\tau} \\
\fl \delta\Psi_1= k^a k^p m^b m^c n^d \nabla_c
C_{p b a d} +2 \Psi_1\beta + \Psi_0\mu - 
 3 \Psi_2\sigma \\
\fl \bar{\delta}\Psi_1= k^a k^p m^b \bar{m}^c n^d \nabla_c
C_{p b a d} -2 \Psi_1\bar{\beta}+ \Psi_0\lambda- 
 3 \Psi_2 \rho + 2 \Psi_1\bar{\tau} 
 \end{eqnarray}
 \begin{eqnarray}
\fl \delta\delta\Psi_0=k^p  k^q  m^a m^b  m^c  m^d \nabla_d\nabla_c
C_{paqb} -24 \Psi_0 \beta^2 + 
 32 \Psi_1 \beta \sigma+ 
 4 \Psi_0\gamma \sigma+ 
 4 \Psi_0\mu \sigma \cr
\myfl - 12 \Psi_2\sigma^2 + 
 4 \Psi_0 \beta \tau- 
 8 \Psi_1 \sigma \tau + \bar{\lambda} D\Psi_0 - \sigma \Delta\Psi_0 + 10\beta\delta\Psi_0 - \tau
\delta\Psi_0 - 8 \sigma \delta\Psi_1 \cr
\myfl + 4\Psi_0 \delta\beta - 4\Psi_1\delta\sigma \\
\fl \delta\bar{\delta}\Psi_0=k^p k^q   m^a  m^b m^c  \bar{m}^d \nabla_c\nabla_d C_{paqb} +8 \Psi_0\beta\bar{\beta} + 
 8 \Psi_1\beta\rho + 4 \Psi_0\gamma\rho - 
 8 \Psi_1\beta\bar{\sigma} + 4\Psi_0\lambda\sigma \cr
 \myfl - 12 \Psi_2\rho\sigma + 4 \Psi_0\bar{\beta}\tau - 
 8 \Psi_0\beta\bar{\tau} + 8 \Psi_1\sigma\bar{\tau} - 4 \Psi_0\tau\bar{\tau} + \mu D\Psi_0 - \rho\Delta\Psi_0 - 4 \bar{\beta}\delta\Psi_0 \cr
 \myfl + 4 \bar{\tau}\delta\Psi_0 - 4 \rho\delta\Psi_1 - 4 \Psi_0\delta\bar{\beta} - 4 \Psi_1\delta\rho + 4 \Psi_0\delta\bar{\tau} + 2 \beta\bar{\delta}\Psi_0 + \tau\bar{\delta}\Psi_0 - 4 \sigma\bar{\delta}\Psi_1 \\
\fl \delta\delta\bar{\Psi}_0= k^p k^q m^a m^b  \bar{m}^c   \bar{m}^d \nabla_b\nabla_a C_{pcqd} -8 \Psi_0\bar{\beta}^2 - 16 \bar{\Psi}_1\beta\rho + 4\bar{\Psi}_0 \bar{\lambda}\rho - 12 \bar{\Psi}_2\rho^2 \cr
 \myfl + 4 \bar{\Psi}_0 \bar{\gamma}\sigma + 20 \bar{\Psi}_0\beta\tau + 20 \bar{\Psi}_1\rho\tau -12\bar{\Psi}_0\tau^2 - 4 \bar{\Psi}_1\sigma \bar{\tau} + \bar{\lambda} D\bar{\Psi}_0 - \sigma\Delta\bar{\Psi}_0 - 6 \beta\delta\bar{\Psi}_0 \cr
 \myfl + 7 \tau\delta\bar{\Psi}_0 - 8 \rho\delta\bar{\Psi}_1 - 4 \bar{\Psi}_0\delta\beta - 4  \bar{\Psi}_1\delta\rho + 4  \bar{\Psi}_0\delta\tau \\
 \fl \delta\bar{\delta}\bar{\Psi}_0= k^p k^q m^a \bar{m}^b \bar{m}^c \bar{m}^d \nabla_a\nabla_d W_{pbqc} + 24 \bar{\Psi}_0 \beta \bar{\beta} + 8 \bar{\Psi}_1 \bar{\beta}  \rho + 4 \bar{\Psi}_0 \bar{\gamma} \rho + 4 \bar{\Psi}_0 \bar{\mu}\rho \cr
 \myfl - 24 \bar{\Psi}_1 \beta\bar{\sigma} - 12 \bar{\Psi}_2 \rho\bar{\sigma} - 20 \bar{\Psi}_0 \bar{\beta} \tau + 20 \bar{\Psi}_1 \bar{\sigma} \tau - 4 \bar{\Psi}_1 \rho\bar{\tau} + \mu D\bar{\Psi}_0 - \rho\Delta\bar{\Psi}_0 \cr
 \myfl + 4 \bar{\beta} \delta\bar{\Psi}_0 - 4 \bar{\sigma} \delta\bar{\Psi}_1 + 4 \bar{\Psi}_0 \delta\bar{\beta} - 4 \bar{\Psi}_1 \delta\sigma - 6 \beta\bar{\delta}\bar{\Psi}_0 + 5 \tau\bar{\delta}\bar{\Psi}_0 - 4 \rho\bar{\delta}\bar{\Psi}_1
 \end{eqnarray}
 \begin{eqnarray}
 \fl D\Phi_{00}= \frac{1}{2}  k^a k^b k^p \nabla_b R_{pa} \\
 \fl \Delta\Phi_{00}=\frac{1}{2}  k^a k^p n^b \nabla_b R_{pa} +2 \Phi_{00} \gamma + 
 2 \Phi_{00}\bar{\gamma} - 
 2 \bar{\Phi}_{01}\tau - 
 2 \Phi_{01}\bar{\tau} \\
 \fl \delta\Phi_{00}= \frac{1}{2}  k^a k^p m^b \nabla_b R_{pa} - 2\Phi_{01} \rho - 2\bar{\Phi}_{01}  \sigma + 2\Phi_{00} \tau  \\
 \fl \delta \Phi_{01} = \frac{1}{2} k^a m^b m^p \nabla_p R_{ab} +2 \Phi_{01} \beta + \Phi_{00} \bar{\lambda} - \Phi_{02} \rho - 2 \Phi_{11}\sigma \\
 \fl {\delta} \bar{\Phi}_{01} =  \frac{1}{2} k^a  m^b   \bar{m}^p \nabla_b R_{ap} -2 \bar{\Phi}_{01}\beta+ \Phi_{00}\mu - 2 \Phi_{11}\rho - \bar{\Phi}_{02} \sigma + 2 \bar{\Phi}_{01}\tau 
 \end{eqnarray}
 \begin{eqnarray}
\fl \delta \delta\Phi_{00}= \frac{1}{2}  k^p k^q m^a m^b \nabla_b\nabla_a R_{pq} + 8 \Phi_{01}\beta\rho + 2\Phi_{00} \bar{\lambda} \rho - 2 \Phi_{02}\rho^2 + 2 \Phi_{00}\gamma\sigma + 2\Phi_{00}\bar{\gamma}\sigma \cr
\myfl + 2 \Phi_{00}\mu\sigma- 8 \Phi_{11} \rho\sigma - 2 \bar{\Phi}_{02}\sigma^2 - 4 \Phi_{00}\beta \tau + 2 \Phi_{01}\rho\tau + 4 \bar{\Phi}_{01}\sigma\tau - 2 \Phi_{00}\tau^2 \cr
 \myfl - 2 \Phi_{01}\sigma\bar{\tau} +\bar{\lambda}D\Phi_{00} - \sigma\Delta\Phi_{00}+ 2 \beta\delta\Phi_{00} + 3 \tau\delta\Phi_{00} - 4 \rho\delta\Phi_{01} - 4 \sigma\delta\bar{\Phi}_{01} \cr
 \myfl - 2 \Phi_{01}\delta\rho - 2 \bar{\Phi}_{01}\delta\sigma+ 2 \Phi_{00}\delta\tau \\
\fl \delta\bar{\delta}\Phi_{00}=  \frac{1}{2}  k^p k^q m^a \bar{m}^b \nabla_a\nabla_b R_{pq} -4\bar{\Phi}_{01} \beta\rho - 4\Phi_{01} \bar{\beta}\rho + 2 \Phi_{00} \gamma\rho + 2 \Phi_{00}\bar{\gamma}\rho + 2 \Phi_{00} \bar{\mu}\rho \cr
\myfl - 4 \Phi_{11}\rho^2 + 4 \bar{\Phi}_{01} \bar{\beta}\sigma + 2 \Phi_{00}\lambda\sigma - 2 \bar{\Phi}_{02} \rho\sigma - 4 \Phi_{01} \beta\bar{\sigma} - 2 \Phi_{02} \rho\bar{\sigma} - 4 \Phi_{11} \sigma\bar{\sigma} \cr
 \myfl + 4 \bar{\Phi}_{01}\rho\tau + 6 \Phi_{01} \bar{\sigma}\tau + 4 \Phi_{00} \beta\bar{\tau} + 2 \Phi_{01} \rho\bar{\tau} - 6 \Phi_{00} \tau\bar{\tau} + \mu D\Phi_{00} - \rho\Delta\Phi_{00} + 2 \bar{\tau}\delta\Phi_{00} \cr
 \myfl - 2 \bar{\sigma}\delta \Phi_{01} - 2 \rho\delta\bar{\Phi}_{01} - 2 \bar{\Phi}_{01} \delta\rho - 2 \Phi_{01}\delta\bar{\sigma} + 2 \Phi_{00}\delta\bar{\tau} - 2 \beta\bar{\delta}\Phi_{00} + 3 \tau\bar{\delta}\Phi_{00}\cr 
 \myfl - 2 \rho\bar{\delta}\Phi_{01}  - 2 \sigma\bar{\delta}\bar{\Phi}_{01} \,.
\end{eqnarray}

\section{NP scalars in flat space}\label{App:Flat}

Let us consider a flat spacetime with a spherical coordinates system $r_\flat,\theta_\flat,\phi_\flat$ centered on the source, hence the static observer velocities are $u^a = \delta_0^a$. The null tetrad is decomposed as
\begin{equation}\label{Defflattetrad}
k^a = u^a + e_r^a\,,\quad n^a = \frac{1}{2}(u^a-e_r^a)\,,\quad m^a = \frac{1}{\sqrt{2}}\left(e_\theta^a + \ii e_\phi^a \right)\,.
\end{equation}
Using
\begin{eqnarray}\label{FlatSpherical}
\myfl && m^b \partial_b e_r^a = \frac{1}{r_\flat} m^a\,,\quad m^b \partial_b \bar{m}^a =-\frac{1}{r_\flat} e_r^a-\frac{\cot \theta_\flat}{\sqrt{2}r_\flat}\bar{m}^a\,,\quad m^b\partial_b m^a=\frac{\cot \theta_\flat}{\sqrt{2}r_\flat}m^a\,,\nonumber\\
\myfl &&e_r^b\partial_b e_r^a = e_r^b\partial_b m^a =0\,,
\end{eqnarray}
we find that for a source emanating from the center ($r=0$), $\chi = r_\flat = \affine$ and
\begin{equation}\label{nonvanishing}
\myfl \rho = -\frac{1}{\affine}\,,\quad \mu = -\frac{1}{2\affine}\,,\quad \beta = \frac{\cot\theta_\flat}{2\sqrt{2}\affine}\,,\quad \delta \beta=-\frac{\csc^2\theta_\flat}{4\affine}\,,\quad \bar{\delta} \beta = -\frac{\csc^2\theta_\flat}{4\affine}\,.
\end{equation}
Other scalars vanish due to the high symmetry of a spherical wavefront, and $A_0\propto 1/\affine$.

Furthermore, when we consider a reference geodesic propagating in the equatorial plane ($\theta_\flat=\pi/2$) then $\beta=0$ but $\delta \beta \neq 0$. 
This reflects that we cannot attach a two-dimensional Cartesian basis on the curved wave front. However, in the equatorial plane, the $\gr{e}_\theta$, $\gr{e}_\phi$ basis is the best approximation of a Cartesian basis since $\beta=0$. We therefore choose to set initial conditions with~\eqref{nonvanishing} evaluated in the equatorial plane at $\affine\ll \affine_{\rm L}$, as illustrated in Fig.~\ref{FigNotation}.

\section{Schwarzschild geodesic at first order in $M$}\label{App:SZO1}

The geodesic equation can be solved in powers of $M$~\cite{Li:2024oke}. We collect here the results at first order in $M$ which we used to compare with the full numerical results. We first solve for $s(r)$ using $\int \dd s = \int \dd r/(\dd r/\dd s)$ and~\eqref{Eqdrds} and expanding the second integrand at first order in $M$. We integrate for a fixed $r_{\rm min}$, choosing the offset in the affine parameter such that $s=0$ when $r=r_{\rm min}$. With this convention $s>0$ in the second part the plane, and $s<0$ in the first part. Note that the impact parameter depends on the fixed $r_{\rm min}$ via~\eqref{rminexact}, which at first order reads $b = r_{\rm min} - M + {\cal O}(M^2) $, and this must be used in~\eqref{Eqdrds} when expanding at first order in $M$. Eventually we find
\begin{equation}
s(r) = \mp \sqrt{r^2-r_{\rm min}^2}  \mp M\frac{\sqrt{r^2-r_{\rm min}^2}}{(r+r_{\rm min})} +\mathcal{O}(M^2)
\end{equation}
which is inverted at first order as
\begin{equation}
\myfl r(s) \approx r_0 - M\frac{r_0(1-r_{\rm min}^2/r_0^2)}{r_0+r_{\rm min}}  \hspace{0,3cm}\text{with}\hspace{0,4cm} r_0\equiv\sqrt{s^2+ r^2_{\rm min}}\,.
\end{equation}
Integration of $\dd \varphi/\dd r = (\dd \varphi/\dd s)(\dd s / \dd r)$ leads to
\begin{equation}
\myfl \varphi(r) = \mp\arccos{\frac{r_{\rm min}}{r}} \mp \frac{M}{r_{\rm min}}\frac{(2r +r_{\rm min})\sqrt{1-r_{\rm min}^2/r^2}}{(r+r_{\rm min})}  + \mathcal{O}(M^2)\,,
\end{equation}
from which we recover that the deflection angle is $4 M/r_{\rm min}\simeq 4M/b$.

\section{Jacobi matrix}\label{App:Jacobi}

The deformation rate matrix is defined by
\begin{equation}\label{DefScomponents}
S_{ab} = - \rho (m_a \bar{m}_b + m_b \bar{m}_a) - \sigma \bar{m}_a \bar{m}_b - \bar{\sigma} m_a m_b\,.
\end{equation}
The Jacobi matrix, which describes the size of a bundle emanating from the source, is related to the deformation matrix via
\begin{equation}\label{DefSab}
S_{ab} = D J_{ac} \cdot J^{-1}_{cb}\,.
\end{equation}
The equations~\eqref{DrhoDsigma} are equivalently formulated as
\begin{equation}\label{JacobiEquation}
D S_{ab} + S_{ac} S_{cb} = {\cal R}_{ab}\quad \Leftrightarrow \quad D^2 J_{ab} = {\cal R}_{ac} J_{cb}\,,
\end{equation}
where the last expression is nothing but the geodesic deviation equation with
\begin{equation}
\myfl {\cal R}_{ab} \equiv R_{cabd}k^c k^d = - \Phi_{00}(m_a \bar{m}_b + m_b \bar{m}_a) - \Psi_0\bar{m}_a \bar{m}_b - \bar{\Psi}_0 m_a m_b\,.
\end{equation}

The angular distance $\chi$, defined from the source, is defined by $\chi^2 \equiv {\rm det} J_{ab}$. From the matrix identity
\begin{equation}
D( {\rm det} J_{ab}) = {\rm Tr}(S_{ab}){\rm det} J_{ab} = (-2 \rho){\rm det} J_{ab}\,,
\end{equation}
we deduce from~\eqref{DefSab} its evolution~\eqref{Dchi}. Since close to the source $J_{ab} \simeq s \delta_{ab}$ and $\rho \simeq -1/s$, then $\chi \simeq s$. It proves convenient to define an amplification matrix ${\cal A}_{ab}(s) \equiv J_{ab}(s)/s$. Let us parameterize it by
\begin{equation}
\myfl {\cal A}_{ab} = (1- {\cal A}^\kappa) (m_a \bar{m}_b + m_b \bar{m}_a) + {\cal A}^\gamma \bar{m}_a \bar{m}_b + \bar{\cal A}^\gamma m_a m_b + {\cal A}^\omega \epsilon_{ab}\,,
\end{equation}
with the screen space Levi-Civita tensor $\epsilon_{ab} \equiv 2 \ii m_{[a} \bar{m}_{b]}$. The Jacobi equation~\eqref{JacobiEquation} translates into a set of coupled differential equations~\cite{Pitrou:2012ge}
\begin{eqnarray}
\left(D^2 + \frac{2}{s}D + \Phi_{00}\right) {\cal A}^\kappa &=& \Phi_{00}+\Re(\Psi_0 \bar{\cal A}^\gamma)\,,\nonumber\\
\left(D^2 + \frac{2}{s}D + \Phi_{00}\right) {\cal A}^\gamma &=& -\Psi_0 (1-{\cal A}^\kappa + \ii {\cal A}^\omega)\,,\label{EqJacobiComponent}\\
\left(D^2 + \frac{2}{s}D + \Phi_{00}\right) {\cal A}^\omega &=&\Im(\Psi_0\bar{\cal A}^\gamma)\,.\nonumber
\end{eqnarray}
In order to solve for $\rho$ and $\sigma$ we first solve for the previous set of equations along the reference geodesic, starting from the source with initial conditions 
\begin{equation}\label{CIJacob}
\left.{\cal A}^{\kappa,\gamma,\omega}\right|_{s=0}=0\,,\qquad \left.D{\cal A}^{\kappa,\gamma,\omega}\right|_{s=0}=0.
\end{equation}
This allows to get the amplification matrix and then the Jacobi matrix. The deformation rate matrix is obtained via~\eqref{DefSab}, eventually allowing to obtain its components from~\eqref{DefScomponents}. 
This is more stable numerically than solving for $\rho$ and $\sigma$ directly with~\eqref{DrhoDsigma} starting at a small but finite $s$.
Finally the angular diameter distance from the source is obtained from 
\begin{equation}
\chi^2 \equiv s^2 {\rm det}({\cal A}_{ab}) = s^2[(1-{\cal A}^\kappa)^2-{\cal A}^\gamma \bar{\cal A}^\gamma + ({\cal A}^\omega)^2 ]\,,
\end{equation}
and the geometric optics amplification is $\mu_{\rm GO} \equiv (s/\chi)^2$.

\section*{References}

\bibliographystyle{unsrt}
\bibliography{biblio}

\end{document}